\documentclass[tikz,11pt,english,a4paper]{article}
\usepackage{float}
\usepackage{jcappub}
\usepackage[utf8]{inputenc}

\usepackage{bm}
\usepackage{algorithm}
\usepackage[noend]{algpseudocode}
\makeatletter
\def\BState{\State\hskip-\ALG@thistlm}
\makeatother

\usepackage{fontawesome5}
\usepackage{hyperref}
\makeatletter
\newcommand{\github}[1]{%
   \href{#1}{\faGithub}%
}
\makeatother

\usepackage{lmodern}
\newcommand{\connect}{\textsc{connect}}
\newcommand{\class}{\textsc{class}}
\newcommand{\montepython}{\textsc{MontePython}}
\newcommand{\tf}{TensorFlow}
\newcommand{\cobaya}{Cobaya}
\usepackage{siunitx}
\DeclareSIUnit \parsec {pc}

\usepackage[T1]{fontenc}
\newcommand{\startus}{\catcode`_ 12\relax}
\newcommand{\stopus}{\catcode`_ 8\relax}

\newcommand{\CC}{C\nolinebreak\hspace{-.05em}\raisebox{.4ex}{\tiny\bf +}\nolinebreak\hspace{-.10em}\raisebox{.4ex}{\tiny\bf +}}
\newcommand{\CLASSpp}{\textsc{class}\nolinebreak\hspace{-.05em}\raisebox{.4ex}{\tiny\bf +}\nolinebreak\hspace{-.10em}\raisebox{.4ex}{\tiny\bf +}}

\newcommand{\annotatefile}[1]{ $\leftarrow$ {\rm #1.}}

\usepackage[edges]{forest}
\definecolor{folderbg}{RGB}{124,166,198}
\definecolor{folderborder}{RGB}{110,144,169}
\newlength\Size
\tikzset{%
  folder/.pic={%
    \filldraw [draw=folderborder, top color=folderbg!50, bottom color=folderbg] (-1.05*\Size,0.2\Size+5pt) rectangle ++(.75*\Size,-0.2\Size-5pt);
    \filldraw [draw=folderborder, top color=folderbg!50, bottom color=folderbg] (-1.15*\Size,-\Size) rectangle (1.15*\Size,\Size);
  },
  file/.pic={%
    \filldraw [draw=folderborder, top color=folderbg!5, bottom color=folderbg!10] (-\Size,.4*\Size+5pt) coordinate (a) |- (\Size,-1.2*\Size) coordinate (b) -- ++(0,1.6*\Size) coordinate (c) -- ++(-5pt,5pt) coordinate (d) -- cycle (d) |- (c) ;
  },
}
\forestset{%
  declare autowrapped toks={pic me}{},
  declare boolean register={pic root},
  pic root=0,
  pic dir tree/.style={%
    for tree={%
      folder,
      font=\ttfamily,
      grow'=0,
    },
    before typesetting nodes={%
      for tree={%
        edge label+/.option={pic me},
      },
      if pic root={
        tikz+={
          \pic at ([xshift=\Size].west) {folder};
        },
        align={l}
      }{},
    },
  },
  pic me set/.code n args=2{%
    \forestset{%
      #1/.style={%
        inner xsep=2\Size,
        pic me={pic {#2}},
      }
    }
  },
  pic me set={directory}{folder},
  pic me set={file}{file},
}

\begin{document}


\title{CONNECT: A neural network based framework for emulating cosmological observables and cosmological parameter inference}

\author[a]{Andreas Nygaard,}
\author[a]{Emil Brinch Holm,}
\author[a]{Steen Hannestad,}
\author[a]{and Thomas Tram}

\affiliation[a]{Department of Physics and Astronomy, Aarhus University,
 DK-8000 Aarhus C, Denmark}

\emailAdd{andreas@phys.au.dk}
\emailAdd{ebholm@phys.au.dk}
\emailAdd{steen@phys.au.dk}
\emailAdd{thomas.tram@phys.au.dk}

\abstract{Bayesian parameter inference is an essential tool in modern cosmology, and typically requires the calculation of $10^5$--$10^6$ theoretical models for each inference of model parameters for a given dataset combination. Computing these models by solving the linearised Einstein--Boltzmann system usually takes tens of CPU core-seconds per model, making the entire process very computationally expensive.
	
In this paper we present \connect{}, a neural network framework emulating \class{} computations as an easy-to-use plug-in for the popular sampler \montepython{}. \connect{} uses an iteratively trained neural network which emulates the observables usually computed by \class{}. The training data is generated using \class{}, but using a novel algorithm for generating favourable points in parameter space for training data, the required number of \class{}-evaluations can be reduced by two orders of magnitude compared to a traditional inference run. Once \connect{} has been trained for a given model, no additional training is required for different dataset combinations, making \connect{} many orders of magnitude faster than \class{} (and making the inference process entirely dominated by the speed of the likelihood calculation). 
	
For the models investigated in this paper we find that cosmological parameter inference run with \connect{} produces posteriors which differ from the posteriors derived using \class{} by typically less than 0.01--0.1 standard deviations for all parameters. We also stress that the training data can be produced in parallel, making efficient use of all available compute resources. The \connect{} code is publicly available for download on GitHub \github{https://github.com/AarhusCosmology/connect_public}.}

\maketitle

\section{Introduction}

For the past two decades the method of choice for cosmological parameter estimation has been based on stochastic optimisation techniques, typically Markov-chain Monte Carlo (MCMC) methods. These methods have the advantage that they are very robust and do not require derivatives of the cost (likelihood) function. They also easily scale to large numbers of parameters which allows for a simple treatment of nuisance parameters. However, a major disadvantage is that a single calculation of the cost function in cosmology can be very expensive because it requires a full solution of the Einstein--Boltzmann equations of linear perturbation theory (and perhaps even a calculation of non-linear corrections). Such a computation typically takes tens of seconds on a single CPU core, and does not parallelise well beyond 10 cores. A fully converged MCMC run, typically requires between $10^5$ and $10^6$ solutions of the Einstein--Boltzmann solver, so the total computation time can easily reach days or weeks, in particular for more complex cosmological models. Furthermore, a new MCMC run must be performed either when a new cosmological model is required or when new data is added. In the latter case, which is common for modern application, the complete analysis with several datasets can be prohibitively expensive numerically.

The purpose of the present paper is to remedy this through a new framework for emulating cosmological observables based on machine learning via neural networks (NN) which we call  \connect{} (\textbf{Co}smological \textbf{N}eural \textbf{N}etwork \textbf{E}mulator of \textbf{C}\textsc{lass} using \textbf{T}\textsc{ensorFlow}). We demonstrate, via a new plug-in written for the publicly available \montepython{} MCMC code~\cite{Audren:2012wb,Brinckmann:2018cvx} that we can reduce the time required for a full MCMC run to hours rather than days or weeks. A similar plug-in for the code \cobaya~\cite{cobaya} has also been implemented. \connect{} assumes a cosmological model but is independent of the targeted dataset, and separates itself from other Einstein--Boltzmann emulators by allowing for user-friendly plug-and-play generation of a neural network emulator for any cosmology that the user may want to investigate, the only requirement being a working \class{} implementation. With a simple Boolean input argument supplied, we have modified \montepython{} to automatically generate training data with \class{}, train a neural network emulator to sufficient precision and conduct the MCMC analysis using this emulator, with a very significant decrease in total computation time relative to simply running an MCMC analysis directly with \class{}.

The idea of using machine learning and specifically neural networks to speed up computations in cosmology has existed for several years. Much focus has been on emulating $N$-body codes (e.g.~\cite{Euclid:2020rfv}) due to them being massively time consuming. Since the training data is expensive to generate, the field of emulating $N$-body codes is markedly data starved. Accordingly, the machine learning tools usually employed in that context include Gaussian processes~\cite{Mootoovaloo_2022, Ho:2021tem} and polynomial chaos expansion coupled with principal component analysis~\cite{Euclid:2020rfv}. However, when large data samples are available, and especially when the dimensionality is large, neural networks are often superior to other supervised learning strategies (as evident, for example, in the recent dominance of neural networks in the ImageNet Large Scale Visual Recognition Challenge~\cite{deng2009imagenet}), and are therefore the obvious choice of strategy for emulating Einstein--Boltzmann codes which are many orders of magnitude faster at generating data than $N$-body codes.

Examples of use of neural networks in emulation of Einstein--Boltzmann codes date back to the early \textsc{CosmoNet}~\cite{Auld:2006pm, Auld:2007qz}, and the approach has since been revisited on numerous occasions with various target variables. \textsc{CLASSNET}~\cite{Gunther:2022pto, Albers:2019rzt} is embedded in \class{} and learns the source functions, reducing the time required to solve linear perturbation equations. Ref.~\cite{Yus_2019} targets LSS angular power spectra, whereas ref.~\cite{Arico:2021izc} learns the linear matter power spectrum, both using neural networks. More recently, \textsc{CosmoPower}~\cite{SpurioMancini:2021ppk} emulates \class{} computations of CMB spectra with temperature, polarisation and lensing anisotropies, as well as the matter power spectrum. \connect{}, contrary to these works, emulates a wide range of customisable outputs: The user simply defines the desired \class{} output variables in an input file, and the \connect{} framework automatically generates a network emulating these. 

Although Einstein--Boltzmann solvers generate data much faster than their $N$-body siblings, the total time required for the combined process of gathering data, training a network from it and performing parameter inference with the network, is still dominated by the data generation. To optimise the emulation scheme, it is therefore most vital to improve on the data gathering method, e.g. by optimising the amount of information extracted from each Einstein--Boltzmann computation or by generating training data in the most important points of the cosmological parameter space. These topics fall under the machine learning field of active learning~\cite{active-learning2, settles2009active, active-learning1}. Most early active learning algorithms focused on selecting new data at regions of large uncertainty of the emulator in so-called uncertainty sampling. Individual active learning algorithms in uncertainty sampling typically differ on how they approximate the uncertainty of the emulator. Query-by-committee~\cite{settles2009active} algorithms estimate the uncertainty as the spread in predictions from a set of learners trained on the currently available dataset, whereas expected model change approaches~\cite{settles2009active}, such as the expected gradient length algorithm~\cite{settles2008}, select new data that optimise an approximate expected improvement of the emulator; e.g. where the new training gradient has the largest magnitude in the case of expected gradient length sampling. In the case of a fully-connected neural network emulator, however, a measure of network uncertainty is not readily available. Neural network uncertainties can be naturally estimated using architectures such as Monte Carlo dropout~\cite{Gal2016} or Bayesian neural networks~\cite{Goan_2020}, but given the scope of the paper, we leave such endeavours for future investigation. 

Furthermore, an important distinction between the classical active learning applications and the one at hand is that in addition to minimising the global emulator uncertainty, we are especially interested in minimising the error in the regions of parameter space that correspond to cosmologies of large likelihood. This duality, i.e. selecting data where (i) the likelihood value is large and (ii) the current emulator uncertainty is large, has been explored previously in the context of cosmological inference. Particular examples of such active learning strategies in cosmology include ref.~\cite{Rogers:2018smb}, in which a Gaussian process is used to emulate the likelihood function, from which new data can be selected based on their weighting according to some balance of the Gaussian process uncertainty and its current estimate of the likelihood at the proposed points. A similar approach was adopted in ref.~\cite{ibanez} in an iterative fashion, as well as in ref.~\cite{Gammal:2022eob}, where it was found that the exploratory behaviour is increasingly important in many-dimensional problems. However, with Einstein--Boltzmann emulators, the overhead introduced by the Gaussian processes in the Bayesian optimisation may cancel this gain in efficiency since Gaussian processes are known to scale disadvantageously with the size of the training data~\cite{Liu_2018}. Additionally, batch acquisition can be non-trivial, and the optimization of the acquisition function itself contributes considerable overhead, rendering such more advanced methods of active learning useful when the generation of data is slow, e.g. for N-body simulations. Indeed, it was shown in~\cite{Gammal:2022eob} that the overhead involved in these acquisition methods often becomes the computational bottleneck when emulating the relatively fast Einstein--Boltzmann emulators.

In this work, we present an iterative data generating procedure that with little overhead combines the focus of data generation around regions of large likelihood while still being spread to reduce uncertainty far from the maximum likelihood. Our algorithm produces parameter space samples with an MCMC chain run by a neural network iteratively trained on the same points, including a method of protecting against spuriously oversampled regions. With this, we find vastly increased emulator accuracy relative to a standard Latin hypercube sampling of training data. We developed the framework with the notoriously difficult posterior of decaying cold dark matter~\cite{Nygaard:2020sow} as a reference, and tested it \textit{blindly} on a $\Lambda$CDM model with variable neutrino mass and degeneracy parameter, on both of which it performs excellently.

This paper is structured as follows. In section~\ref{sec:2} we specify the design of the neural network architecture employed in \connect{}, and in section~\ref{sec:sample} we describe the novel iterative algorithm for placing training data at advantageous points in the parameter space. In section~\ref{sec:mp} we describe the use of \connect{} through \montepython{} and present resulting MCMC analyses, using \connect{}, for the decaying cold dark matter and massive neutrino cosmological models. Finally, we discuss and conclude on our findings in section~\ref{sec:conclusion}.


\section{Neural network design} \label{sec:2}

The method used for the emulation is a fully connected deep neural network consisting of an input layer, multiple hidden layers, and an output layer (see e.g.~\cite{ddl} for a recent overview). The input layer consists of the cosmological parameters from which we would like to extract an output, i.e. any numeric parameter that the Einstein--Boltzmann solver code \class{} takes as input~\cite{ddl}. The hidden layers have a much larger dimension of several hundreds of nodes, in order to create enough trainable weights for the network to find the correct behaviour of the \class{} computations. The output layer consists of all the specified spectra and output parameters one wishes to emulate -- this being any output that \class{} can compute (CMB spectra, matter power spectra, background functions, thermodynamical parameters, and derived parameters).

The first step is to gather training data for the network which requires a method for sampling in the space of cosmological input parameters. The construction of this method will be further discussed in section~\ref{sec:sample}. When the sample of input values has been constructed, we can use \class{} to calculate the specified output values for each point in the sampled data. This is then combined to a single output array for each point while the cosmological input parameters are put in a single input array for each point. Together, the set of input arrays and output arrays constitute our training data.


Using the \tf{} framework~\cite{tensorflow} we can now train the network on the training data for a specified number of \emph{epochs}, where an epoch refers to an update of the network weights. Each network used for our results has been trained for 300 epochs\footnote{Except for a single massive neutrino model trained on a Latin hypercube consisting of $10^6$ points used solely for comparison in figure~\ref{fig:mcmc-mas_lhc}. This has only trained for 100 epochs due to the large amount of data causing the optimiser to diverge, but the accuracy of the network stagnates quickly so there would be little to no gain with more epochs anyway.} and with batch-sizes of 512. The \textit{loss function} is then minimised using a specified optimisation algorithm (We have used the ADAM optimiser~\cite{Adam} for this work and as the default in \connect{}) which slightly tweaks the weights of the network while propagating backwards. The Network is then ready for the next epoch where the whole procedure is repeated in order for the network to perform better with each epoch.

\subsection{Network architecture}

An Einstein--Boltzmann solver can be seen as a function mapping the cosmological parameters into a set of observables such as the CMB anisotropy coefficients or the linear matter power spectrum. Since this mapping can be very general, the most conservative neural network structure to employ is a fully connected, feed-forward, deep neural network~\cite{ddl}\footnote{Other Einstein--Boltzmann emulators have used different architectures. For example, ref.~\cite{Albers:2019rzt} used convolutional layers~\cite{ddl}. However, such choices are always motivated by some properties of the underlying physics, and since \connect{} emulates customizable \class{} outputs, we cannot directly make such assumptions.}. This involves several hidden layers where each node in a layer is connected to each node in the next layer. For all results in this paper we use 6 hidden layers with 512 nodes in each, inspired by the architecture in ref.~\cite{SpurioMancini:2021ppk}. Too few nodes and layers restrict the ability of the network to emulate the desired computation, and too many both make it prone to over-fitting~\cite{ddl} and require larger datasets. We find that our chosen values evade both of these concerns, and by varying these network parameters slightly, we find only modest changes in the network performance. We therefore leave a more thorough investigation of the optimal network architecture for future work. However, one soft requirement is that the evaluation time of the \connect{} architecture must not exceed the evaluation time of typical likelihood codes such as Planck~\cite{Planck:2019nip} or Planck lite~\cite{Prince:2019hse}. We have conducted rough benchmarking of the likelihood codes and the \connect{} evaluation time, and find that at around 12 layers ($\sim\!3\times10^6$ trainable parameters), the evaluation time of \connect{} becomes greater than the evaluation time of \textit{Planck lite}, giving an approximate upper bound on the architecture complexity. Furthermore, since \connect{} allows the user to choose the outputs to emulate, one should keep in mind that the ideal architecture will vary with the size of the output, with larger outputs naturally requiring a larger network complexity. 
For example, Einstein--Boltzmann solvers such as \class{} do not evaluate $C_{\ell}$ coefficients for each $\ell$, but rather at a reduced set of approximately $10^2$ $\ell$-values from which the full sets of $C_{\ell}$ coefficients are constructed by interpolation.
This significantly reduces the output dimension and we have consequently chosen the set of $\ell$-values directly computed by \class{} for the output layer.

\subsection{Choosing a loss function}

When training a neural network, one always has to make choices regarding the optimisation of the network. First of all, we need a way of quantifying how well the output from the network fits the desired output from the training data -- the \textit{loss function}~\cite{ddl}. A simple choice for a loss function would be the widely-used \textit{mean squared error} (MSE) function,
\begin{equation}\label{eq:mse}
	L_{\rm MSE}({\bm x},{\bm y}) = \frac{1}{n}\sum^{n}_{i=1} (x_i - y_i)^2,
\end{equation}
where ${\bm x}$ is the output from the network and ${\bm y}$ is the output from the training data. This loss function ensures that the network performs equally well on every output node and is thus the apparent choice if we are to remain agnostic about our network.

\begin{figure}[tb]
	\centering
	\includegraphics[width=\textwidth]{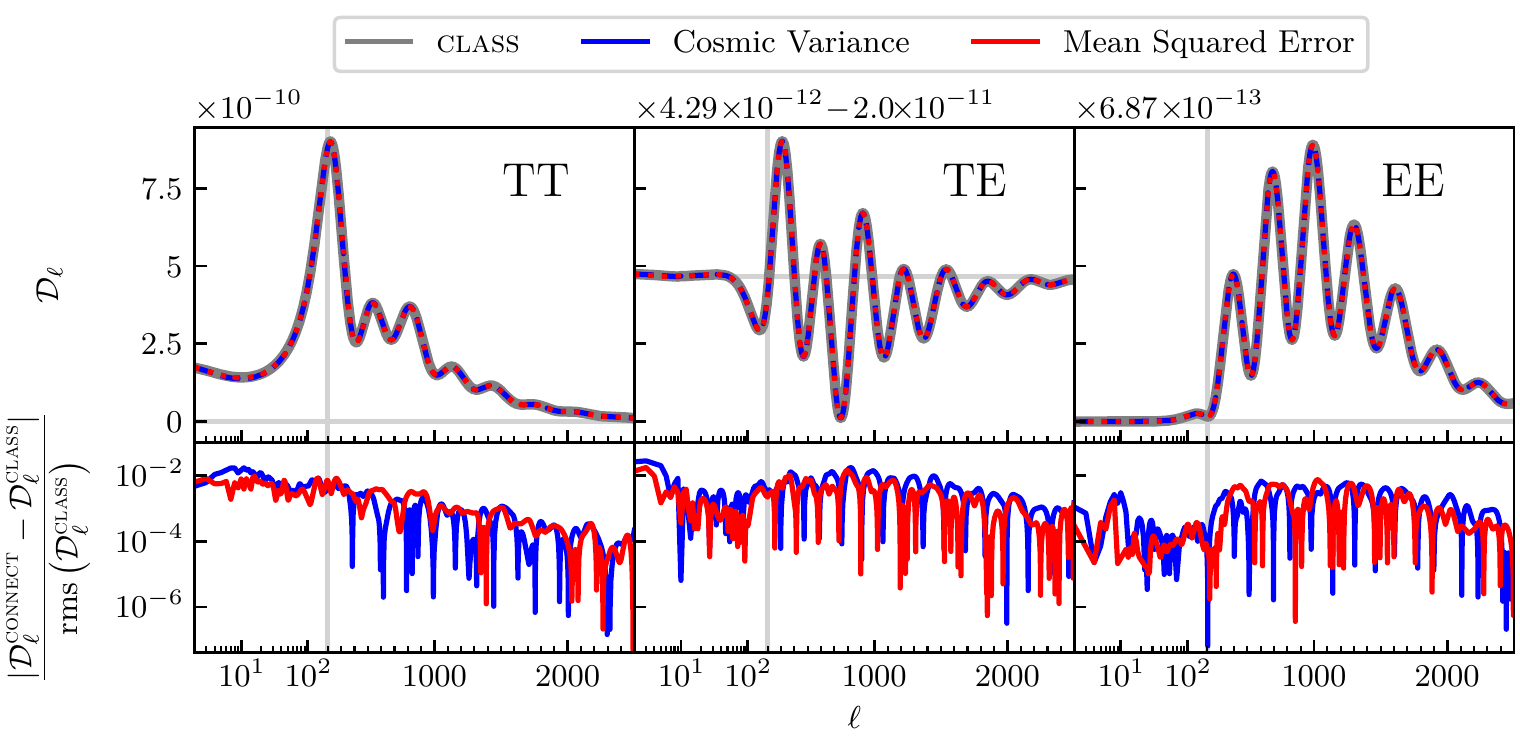}
	\caption{\label{fig:cmb}CMB spectra as calculated by \textsc{class} along with two neural network models with different loss functions. The difference is not really visible in the spectra, so the errors are included as well. Due to some of the spectra values being close to zero, a relative error would be misleading. The absolute errors are therefore scaled by the rms-values of the spectra.}
\end{figure}

There are, however, various situations where this approach is not the most optimal, and the CMB spectra are examples hereof. Measurement errors on the CMB 
temperature and polarisation power spectra are a combination of cosmic (sample) variance, noise, and finite beam width (see e.g.~\cite{Wang:1998gb}). Modern CMB probes, such as Planck, in general provide spectra which are cosmic variance limited (except for $B$-mode polarisation) effectively out to the maximum $\ell$-value measurable with the given beam width, and can therefore be reasonably approximated by assuming cosmic variance out to some maximum $\ell$ beyond which the error goes to infinity. Using this observation as a guide we therefore modify equation~\eqref{eq:mse} with $\ell$-dependent coefficients,
\begin{equation}\label{eq:cv}
	L_{\rm CV}({\bm x},{\bm y}) = \frac{1}{n}\sum^{n}_{i=1} \frac{2\ell_i + 1}{2}(x_i - y_i)^2,
\end{equation}
Figure~\ref{fig:cmb} shows the CMB spectra of a $\Lambda$CDM model as calculated by \class{} and two neural network models with different loss functions, $L_{\rm MSE}$ and $L_{\rm CV}$. The figure also includes the errors between the spectra from the neural network models and \class. The errors are calculated as the absolute difference scaled by the rms-values of the spectra. This is due to the fact that a normal relative error is misleading when the values of the spectra are close to zero, since the error would be very large even though the discrepancy is rather small.

\begin{figure}[tb]
	\centering
	\includegraphics[width=\textwidth]{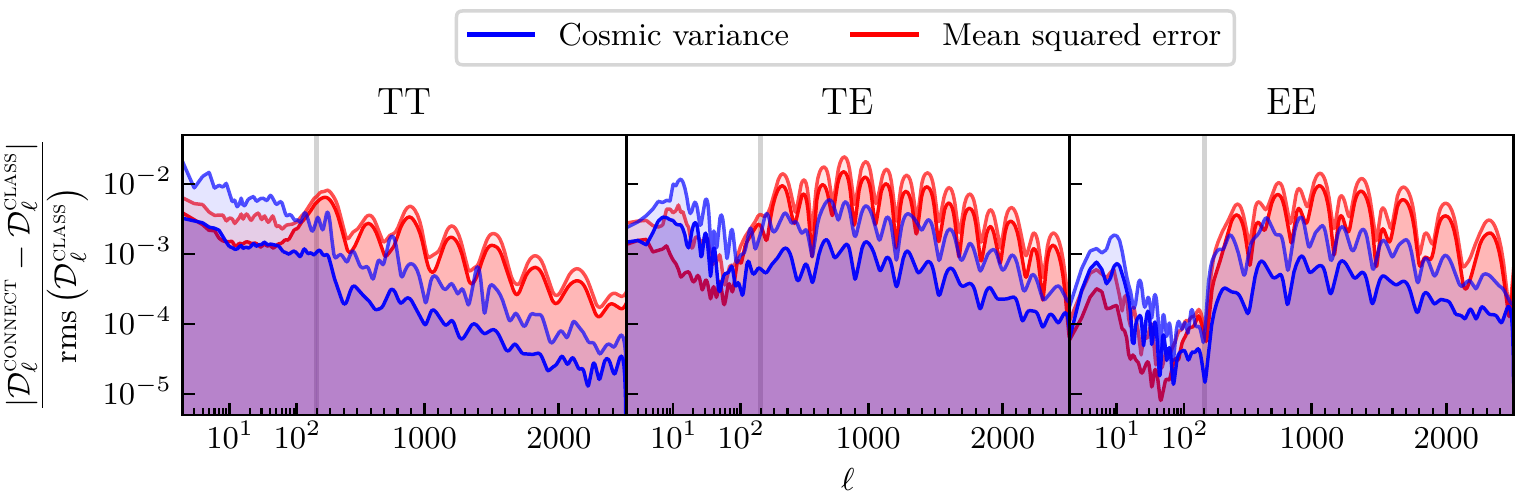}
	\caption{\label{fig:error_loss}Percentiles of errors in the CMB spectra of neural network models with different loss functions when using a test dataset of 20,000 points from a high-temperature MCMC sampling of the DCDM posterior. Both the $1\sigma$ and $2\sigma$ percentiles are included for each model.}
\end{figure}

Evaluating the loss functions on a single cosmological model may misrepresent the performances on a larger region of cosmological parameter space. We therefore use the neural network models on a set of data from a high-temperature MCMC sampling containing 20,000 points and calculate the error in the same way. Figure~\ref{fig:error_loss} shows the $1\sigma$ and $2\sigma$ percentiles of this set of errors for both loss functions.
It is clear from figure~\ref{fig:error_loss} that the cosmic variance loss function has the desired effect of improving the accuracy at high $\ell$s at the cost of accuracy at low $\ell$s. This in turn improves results of parameter inference compared to using the mean squared error loss function. When including additional output, such as derived parameters, we need to use a combination of both loss functions, but we also need to assign an importance to all non-CMB output similar to that of the highest $\ell$-mode, as to not get a low accuracy on these.

\subsection{Choosing an activation function}

\begin{figure}[tb]
	\centering
	\includegraphics[width=\textwidth]{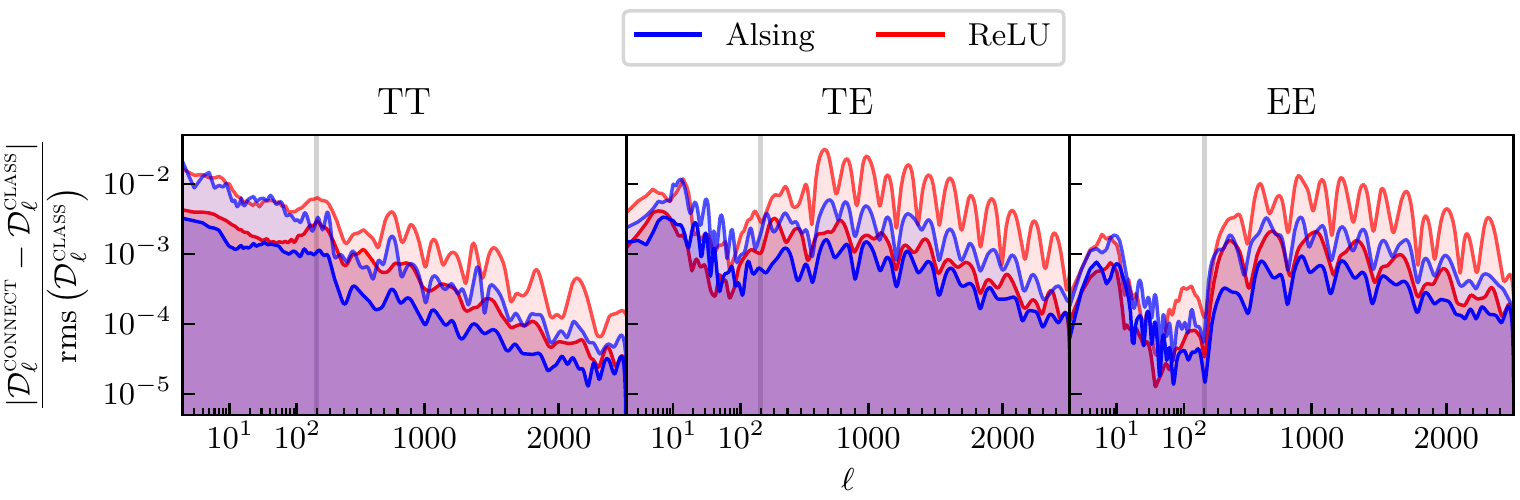}
	\caption{Percentiles of errors in the CMB spectra of neural network models with different activation functions when using a test dataset of 20,000 points from a high-temperature MCMC sampling of the DCDM posterior. Both the $1\sigma$ and $2\sigma$ percentiles are included for each model. "Alsing" refers to equation~\eqref{eq:alsing} as presented in ref.~\cite{Alsing:2020}.}
	\label{fig:error_activation}
\end{figure}

Another choice we need to make is the choice of an \textit{activation function}~\cite{ddl}. The traditional choice is the \emph{Rectified Linear Unit} (ReLU) function~\cite{relu}. However, the main drawback of ReLU is that the training might become more difficult due to the derivative being exactly zero for negative input. For our application, we found that the following parameterised ReLU with a smoothing between the positive and negative parts, as suggested in ref.~\cite{Alsing:2020}, works well. There are two free parameters of this activation function, one for the slope of the negative part and one for the smoothing, and we allow these to be trained alongside the weights of the network. We can furthermore assign different parameters for each node in a layer which will then be optimised during training. This leads to the form of the activation function as presented in ref.~\cite{Alsing:2020},
\begin{equation}\label{eq:alsing}
	{\bm f}({\bm x}) = \left({\bm \gamma} +\left( 1+ {\rm e}^{-{\bm \beta}\odot {\bm x}} \right)^{-1} \odot \left( 1 - {\bm \gamma}\right)\right) \odot {\bm x},
\end{equation}
where the parameters ${\bm \beta}$ and ${\bm \gamma}$ control the smoothing and slope of the negative part, respectively, and the $\odot$ represents elementwise multiplication. From figure~\ref{fig:error_activation}, it is evident that this activation function performs better than the simple ReLU activation function.

\subsection{Normalisation of inputs and outputs}

When using an artificial neural network it is beneficial, and often necessary, to consider scaling of the training data~\cite{ioffe2015}. This is especially true in our case, since the input nodes and output nodes vary with several orders of magnitude. If we were to not consider this at all, the loss function would only have significant contributions from the larger values, while small numbers, such as the $C_{\ell}$s, would have a vanishing impact on the total loss. We are therefore required to address this problem in some manner. There are several ways to deal with this and they include the following:
\begin{enumerate}
	\item \textbf{Min-Max scaling}, where data belonging to each node in the input (output) layer is transformed to the same interval, e.g. $[0,1]$, using the minimal and maximal values of the data within the node: $X_{\rm new} = (X - X_{\rm min})/(X_{\rm max} - X_{\rm min})$.
	
	\item \textbf{Logarithmic scaling}, where the data is transformed to logarithmic space in order for values differing by several orders of magnitude to lie within the same order of magnitude (or few apart). Since $X_1/X_2 = \exp[\log(X_1)-\log(X_2)]$, this also ensures that optimisation of absolute loss in logarithmic space is equivalent to an actual optimisation of relative loss, meaning that larger orders of magnitude will not be favoured above smaller orders of magnitude.
	
	\item \textbf{Standardisation}, where data belonging to each node in the input(output) layer is transformed to a normal distribution with zero mean ($\mu = 0$) and a variance of unity (${\rm Var} = 1$): $X_{\rm new} = (X - \mu)/\sqrt{{\rm Var}}$.

\end{enumerate}
The input arrays in the training data are automatically normalised with standardisation using \tf{}'s own \texttt{preprocessing.Normalization} routine based on a usual batch normalisation scheme~\cite{ioffe2015}. The means and variances are stored as weights in the input layer of the model and we thus do not need to do anything explicit to the inputs. It is not quite as easy with the output arrays, since no similar routine is available for the output layer. We instead have to normalise the output arrays manually, and we have therefore implemented all of the above three methods in \connect{}.

\begin{figure}[tb]
	\centering
	\includegraphics[width=\textwidth]{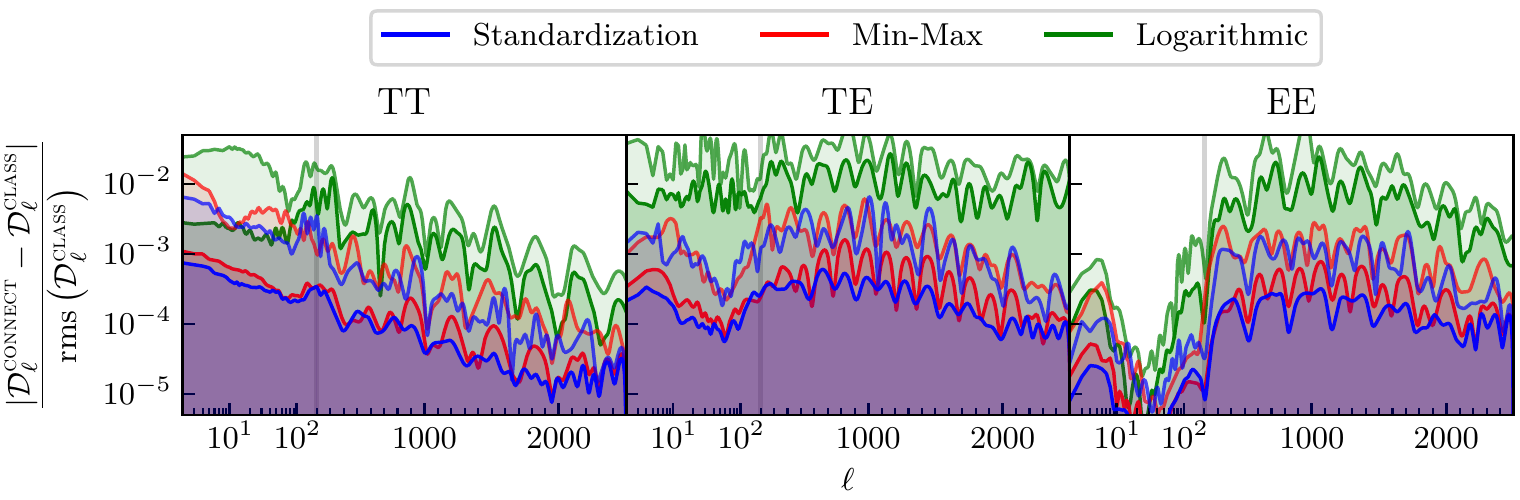}
	\caption{Percentiles of errors in the CMB spectra of neural network models with different normalisation methods when using a test dataset of 20,000 points from a high-temperature MCMC sampling of the DCDM posterior. Both the $1\sigma$ and $2\sigma$ percentiles are included for each model.}
	\label{fig:error_normalization}
\end{figure}

All three methods of normalisation yield good results when compared to simply multiplying all spectra with a constant factor of $10^{10}$, but the accuracy is much better when using min-max scaling or standardisation. This is because all nodes in the output layer have the same span and the network treats them similarly. When using logarithmic scaling, the order of magnitude is still very similar, but there is a clear difference between the $C_{\ell}$s and other kinds of output, such as derived parameters, since the $C_{\ell}$s will have values around or lower than $-20$ while other parameters will have values closer to or above zero. This leads to a difference in how the output nodes are viewed and treated by the network, and it is thus harder to achieve convergence. In our implementation of logarithmic scaling, we found that the performance can be further increased by taking the logarithm twice (after a shift of all the data to positive values) since the difference in orders of magnitude for $C_{\ell}$s and derived parameters is quite large. Standardisation yields a slightly better result than min-max scaling, as apparent from figure~\ref{fig:error_normalization}, and so it has been chosen as the default normalisation method. All results in this paper have been produced with standardisation as the normalisation method except for the comparisons between different methods of normalisation in this section.

\section{Sampling of training data}\label{sec:sample}

The training data can be sampled in various ways with different methods having different strengths and weaknesses. The most agnostic way of sampling the parameter space would be using a grid-based method. To get a good resolution these can, however, be very costly and we end up with many points that yield almost identical output since many of them only differ in a single parameter. To circumvent this, we can use \textit{Latin hypercube sampling}, where no two models share any parameter values. This is much more efficient and proves sufficient for the training of the neural network. This way of sampling still yields a uniform distribution of points in the parameter space, so the trained network will be able to emulate the output for all models in the parameter space (within the boundaries of the Latin hypercube) with similar accuracy.

For a large Latin hypercube containing all reasonable models, it is, however, rare that we would ever need to use the network on the outer parts of the hypercube. This is due to the fact that most models near the edges (and especially the corners) have very low likelihoods since they are very far from the best-fit points of most datasets. If we disregard all such unlikely models, a better way of sampling would be by mimicking the shape of the actual posterior distribution. With high dimensionality in the parameter space, this proves much more efficient than using Latin hypercube sampling, since only a small fraction of the models are of actual use in the latter. A way of illustrating this effect is by imagining a simple hyperspherical posterior with radius $R$ centred around the best-fit point. The ratio of the volume of the hypersphere to that of a hypercube with side length $2R$ is given by
\begin{equation}
	r_n = \frac{V^{\rm sphere}_n}{V^{\rm cube}_n} = \frac{\pi^{n/2}}{2^n\,\Gamma\big(\frac{n}{2}+1\big)},
\end{equation} 
and in high-dimensional space this decreases rapidly. With just 3 parameters, the corners of the hypercube makes up almost half of the volume, and with 9 parameters, less than a percent of the volume is within the hypersphere. By only focussing on models within such a hypersphere, we could utilise our resources much better and increase the performance of the network on all the relevant models of interest. 
Actual posteriors typically have a much more complicated shape than a hypersphere, however, the argument still holds due to many of the cosmological parameters having a vanishing likelihood only a few standard deviations away from the best-fit point. We cannot simply expect that a hyperspherical sampling will be representative of the posterior distribution. We thus need a way of sampling training data from the actual posterior distribution instead. We therefore propose to sample the training data using an MCMC method with a high sampling temperature. It seems a little strange to use an MCMC method to create the training data for a neural network that is to be used in an MCMC analysis, but the idea is that we do not need anywhere near as many data points for the training data as we do for the actual MCMC analysis. A high-temperature MCMC sampling running for a few hours is sufficient to obtain the same (or better) accuracy on the relevant models as one would get from Latin hypercube sampling with $10^5$--$10^6$ points. As noted in ref.~\cite{Schneider_2011}, we thus obtain a set of training data from the exact region of the parameter space where emulation is relevant instead of having the majority of the training data unrealistically far away from the best-fit point.

In this paper, we present results obtained with the default \connect{} temperature of $T=5$. Since the temperature alters the likelihood $\mathcal{L}$ as $\mathcal{L} \rightarrow \mathcal{L}^{1/T}$, a temperature of $T=5$ corresponds to increasing the standard deviation of a Gaussian likelihood by a factor 5, to the effect that the generated training data mainly lies inside the $5\sigma$ contour of the posterior. However, this is a free input parameter and may be adjusted if the user desires higher accuracy further away from the posterior mode.

\subsection{Iterative sampling}

\begin{figure}[tb]
	\centering
	\includegraphics[width=\textwidth]{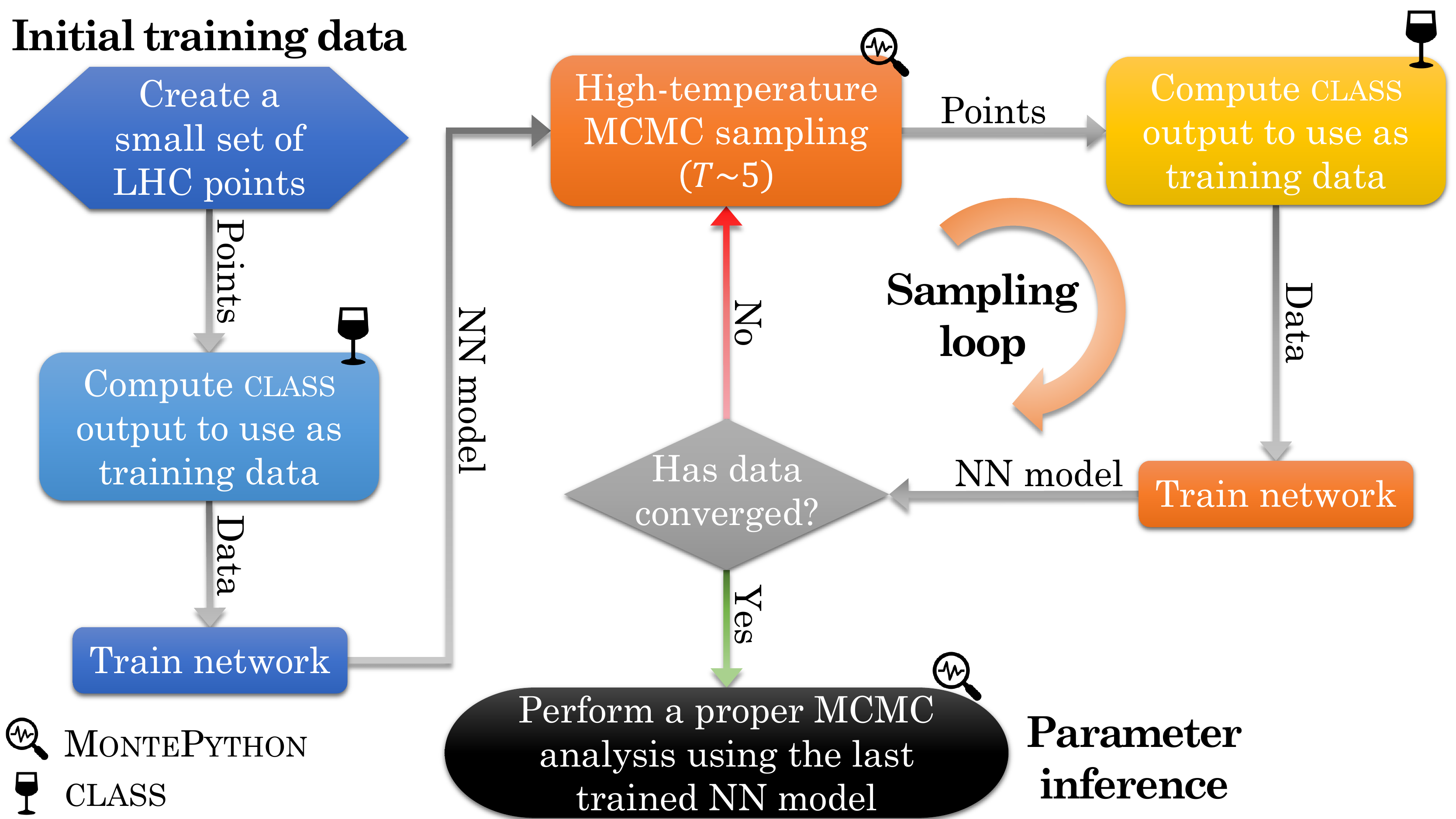}
	\caption{Flowchart of the iterative sampling algorithm.}
	\label{fig:flow}
\end{figure}

We can even improve on this and make the sampling even more efficient. We can exploit the fact that we only need to sample from something roughly similar to the posterior distribution and not the actual one, due to the high sampling temperature and the neural networks ability to interpolate in a well-sampled area. We therefore do not need to search the parameter space with the precision of \class{} resulting in many slow calculations of the likelihood. With a typical acceptance rate of $0.3$ we are wasting a lot of computation time when calculating the likelihood of rejected steps. A way around this is to use a neural network trained with only a small number of Latin hypercube points ($\sim10^4$) -- enough to give a decent, but not great, accuracy. We then use this neural network model to calculate likelihoods during the MCMC sampling much faster and sample new points for the training data. We then only need to use \class{} to calculate the output-part of this new training data, which means that we effectively skip the \class{} computations of all the rejected steps. Since the neural network model had a relatively low accuracy, we have only gained a rough sample around the posterior distribution, but new models trained with this new training data shows a major improvement. We can even repeat the process starting from this new model, and the resulting models improve for each iteration. Using this method we can thus get rid of a huge part of the expensive \class{} computations only resulting in rejected steps. We could include the initial Latin hypercube data in the total dataset, but by doing so we lower the accuracy of the neural network model in the relevant parts of the parameter space. This is due to the fact that the network contains a limited set of weights, and by forcing it to learn the behaviour in the outermost regions of the parameter space, we inhibit the training in the more relevant regions. It is thus advantageous to exclude the initial training-data even though the \class{} computations have already been done. An illustrative flowchart of this sampling algorithm is shown in figure~\ref{fig:flow}. During the completion of this paper a few similar approaches have been published~\cite{LINNA, boruah2022, Wang_2020}, which further increases the confidence in this type of sampling.

When using this way of sampling, we are interested in the least amount of points possible with the best representation of the posterior for a good accuracy. We are therefore not interested in using all of the points from the high-temperature MCMC samplings, since the burn-in period yields unfavourable points to use as training data. In order to get a representative set of training data, we therefore sample longer MCMC chains and keep only the last $N$ points for the training data. The question now remains how to determine when each high-temperature MCMC sampling should end as well as when the accuracy of an iteration is acceptable. We propose similar answers to the two questions, namely to stop when the variance falls below a certain threshold. For the individual high-temperature MCMC samplings it is the variance between the chains, and for the iterations it is the variance between the kept points from two consecutive iterations. 

\subsection{Reduction of over-densities in sample points}
We could use all of the kept points from each MCMC run, but we would then get a lot of similar points in our total dataset, since each iteration roughly samples from the same distribution. It would be beneficial to have a way of determining which points we can safely discard, as to not waste computational power increasing our dataset where it is already well sampled. A simple, yet effective, way of doing this is the following:\\

\begin{algorithmic}
\State $P_i\hspace{10.9pt}=\mathrm{new\;points\;from\;current\;iteration}$
\State $P_{i-1}=\mathrm{points\;in\;the\;data\;set\;from\;previous\;iterations}$

\For{$p\;\textbf{in}\;P_{i}$}

\State $x\hspace{17.1pt}=\mathrm{nearest\;point\;of}\;p\;\mathrm{in\;}P_{i-1}$
\State $d_{\rm min}\hspace{3pt}=\mathrm{distance\;between\;}p\;\mathrm{and\;}x$
\State $D_{i-1}=\mathrm{array\;of\;distances\;between}\;x\;\mathrm{and\;the}\;n\;\mathrm{nearest\;neighbours\;of}\;x\;\mathrm{in\;}P_{i-1}$
\State $D_{i}\hspace{10.9pt}=\mathrm{array\;of\;distances\;between}\;p\;\mathrm{and\;the}\;n\;\mathrm{nearest\;neighbours\;of}\;p\;\mathrm{in\;}P_{i}$

\If {$d_{\rm min} > {\rm average}(D_{i-1}) + 2\cdot{\rm std}({D_{i-1}})$}
\State $p\;\mathrm{is\;accepted}$
\ElsIf {${\rm average}(D_{i}) < {\rm average}(D_{i-1}) - 2\cdot{\rm std}({D_{i-1}})$}
\State $p\;\mathrm{is\;accepted}$
\EndIf

\EndFor
\State $\mathrm{Add\;all\;accepted\;points\;to\;the\;data\;set}$
\end{algorithmic}
\vspace{1em}
The conditions of the if-statements might seem arbitrary at first, but we found that a tolerance of 2 standard deviations gave the most consistent results. There are several reasons why oversampling should be avoided, and computational waste is only one of them. Another reason is that an oversampling of certain regions leads to a bias in the training, since these regions are given more weight in the calculation of the total loss. When repeating the sampling for several iterations, the oversampled regions may differ between two iterations and thus result in trained neural networks with different biases. These will in turn lead to different samples, and the convergence in the data between two consecutive iterations will be immensely difficult and require a large number of iterations. 

\begin{figure}[tb]
	\centering
	\includegraphics[width=\textwidth]{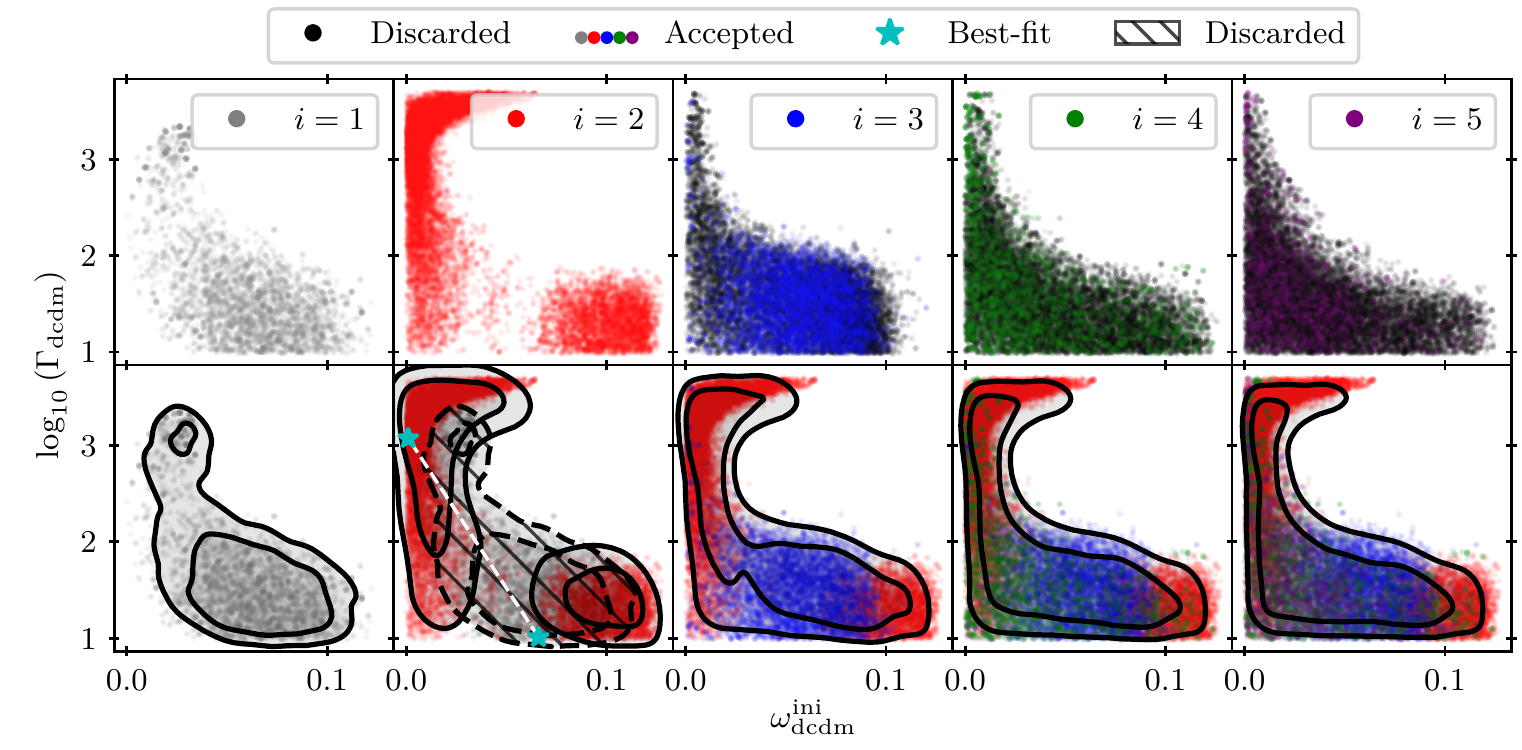}
	\caption{Upper panel: Data from each iteration of a sampling of a decaying cold dark matter cosmological model in the $(\omega_{\rm dcdm}^{\rm ini}, {\rm log}_{10} (\Gamma_{\rm dcdm}))$  plane. Data points from each iteration are filtered to prevent oversampling of certain regions, and the accepted points are shown in color.
	Lower panel: Combined data after filtering including $1\sigma$ and $2\sigma$ contours. Note how the best-fit point of iteration $i=1$ is several standard deviations away from the subsequent best-fit point. Including the $i=1$ data in the final training set would degrade the performance of the network as argued in the text.}
	\label{fig:iterations}
\end{figure}

Next, in figure~\ref{fig:iterations} we provide an actual demonstration of how the iterative procedure works for the case of a decaying cold dark matter (DCDM) model. 
This model is described by a number of cosmological parameters: $\omega^{\rm ini}_{\rm dcdm}$ and $\Gamma_{\rm dcdm}$ (see e.g. ref.~\cite{Nygaard:2020sow} for details on the model parameters).
The figure shows the training data acquired in each iteration in the 2-dimensional $(\omega_{\rm dcdm}^{\rm ini}, {\rm log}_{10} (\Gamma_{\rm dcdm}))$ parameter space, and only the accepted points using the aforementioned algorithm are shown in color.

From the different iterations of figure~\ref{fig:iterations} it is clear that we might need to further discard some of the points in our dataset. The points from the first iteration, as sampled by the neural network model trained on the initial Latin hypercube data, often have little to no overlap with those from the other iterations, and including them in the dataset thus leads to a worse accuracy in the relevant part of the parameter space, since the network has to focus some of the training on an irrelevant region. It would therefore be beneficial to remove the data from the first iteration all together like we removed the initial Latin hypercube data. This way of discarding data is more wasteful than the filtration of points from the MCMC samplings, since we again are throwing away already computed \class{} models. In order to combat this waste of resources, we decrease the number of points sampled by the initial neural network model. For some cosmological models the first sample is not far from the consecutive samples, and in those cases we could keep the data from the first iteration without lowering the accuracy. We have therefore included the option of keeping the data from the first iteration if one wishes to do so. When looking at figure~\ref{fig:iterations}, one could make an argument for keeping the first iteration (or, though wasteful, throwing away the second as well), but we discard it to be on the safe side. The first iteration contains 5,000 \class{} computations and the maximal amount of new points from each iteration is 20,000. The final dataset contains 19,999 points from $i=2$ (one \class{} computation returned an error and was excluded), 7,475 points from $i=3$, 4,781 points from $i=4$, and 2,415 points from $i=5$. We thus see that the amount of points taken from each iteration decreases due to convergence, so less and less \class{} computations need to be performed.

\section{Integration with MontePython}\label{sec:mp}

In order to gain any real benefits of a neural network emulating cosmological observables, we need to be able to use the network instead of an Einstein--Boltzmann solver code like \class{} in an MCMC analysis. We have therefore made a module for \connect{} as a plug-in for the popular MCMC code \montepython{}. Using this plug-in along with the \textit{Planck lite} likelihood \cite{Prince:2019hse}, one can reach speedups of 2--3 orders of magnitude. This means that a reasonable inference can be done in mere minutes.

\subsection{Considerations and usage}

Now that the computation speed of CMB spectra is increased significantly, it no longer dominates the computation time of each step in an MCMC chain. This means that the computation time of the likelihood dominates the computation time when using the \connect{} plug-in, and this means that we are limited to only certain likelihoods if we want the greatest speedups. When using the full \textit{Planck clik} likelihood, we only see speedups of less than one order of magnitude using the \connect{} plug-in, and this is because of the vast number of nuisance parameters making the likelihood computation slow. To really see the benefits of the plug-in, we need to use the much faster \textit{Planck lite} likelihood which is marginalised over the nuisance parameters leaving only a single one for the likelihood computation. This leads to speedups of several orders of magnitude. 

A few things in the source code of \montepython{} are specific to \class{} and for the sake of usability we did not want to alter anything in the source code. The solution was therefore to make our new plug-in inherit from the cython wrapper of \class{}, \texttt{classy}, and trick \montepython{} into believing that our \connect{} module is \class{}. This way, a user will not have to alter any code, and any version of \montepython{} supporting \class{} can be used. 
\startus
One simply has to set the path of the cosmological module to that of the \connect{} plug-in instead of a \class{} installation and add the name of a trained \connect{} model to the \texttt{data.cosmo_arguments} dictionary in the parameter file.
\stopus

Since the plug-in inherits from the \texttt{classy} wrapper, it will automatically use \class{} to compute any derived parameter that was not emulated by \connect. Since the background calculation of \class{} is at the same order of magnitude of computation time as a \connect{} emulation, we can just let \class{} take care of any derived parameter that is only dependent of the background module of \class. It is, however, a good idea to include any other derived parameter, needed for the MCMC analysis, in the emulation output, since all other modules of \class{} are slower and this would impact the computation time significantly.

\subsection{Inference with Planck lite}


\begin{figure}[tb]
	\centering
	\includegraphics[width=\textwidth]{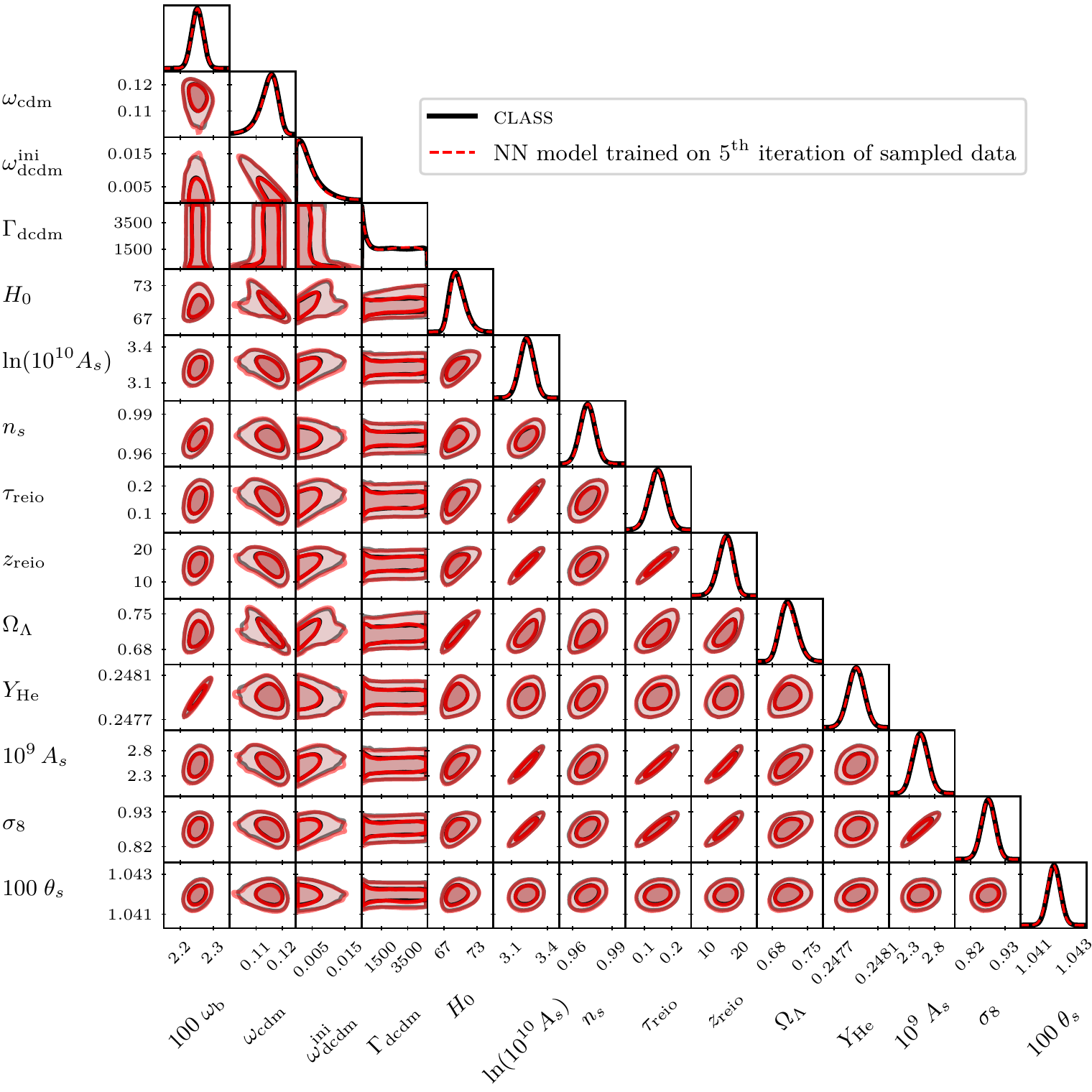}
	\caption{1D and 2D posteriors for the DCDM model resulting from a standard \class{}-based run (black) and \connect{} (red).}
	\label{fig:mcmc-dcdm}
\end{figure}


\begin{figure}[tb]
	\centering
	\includegraphics[width=\textwidth]{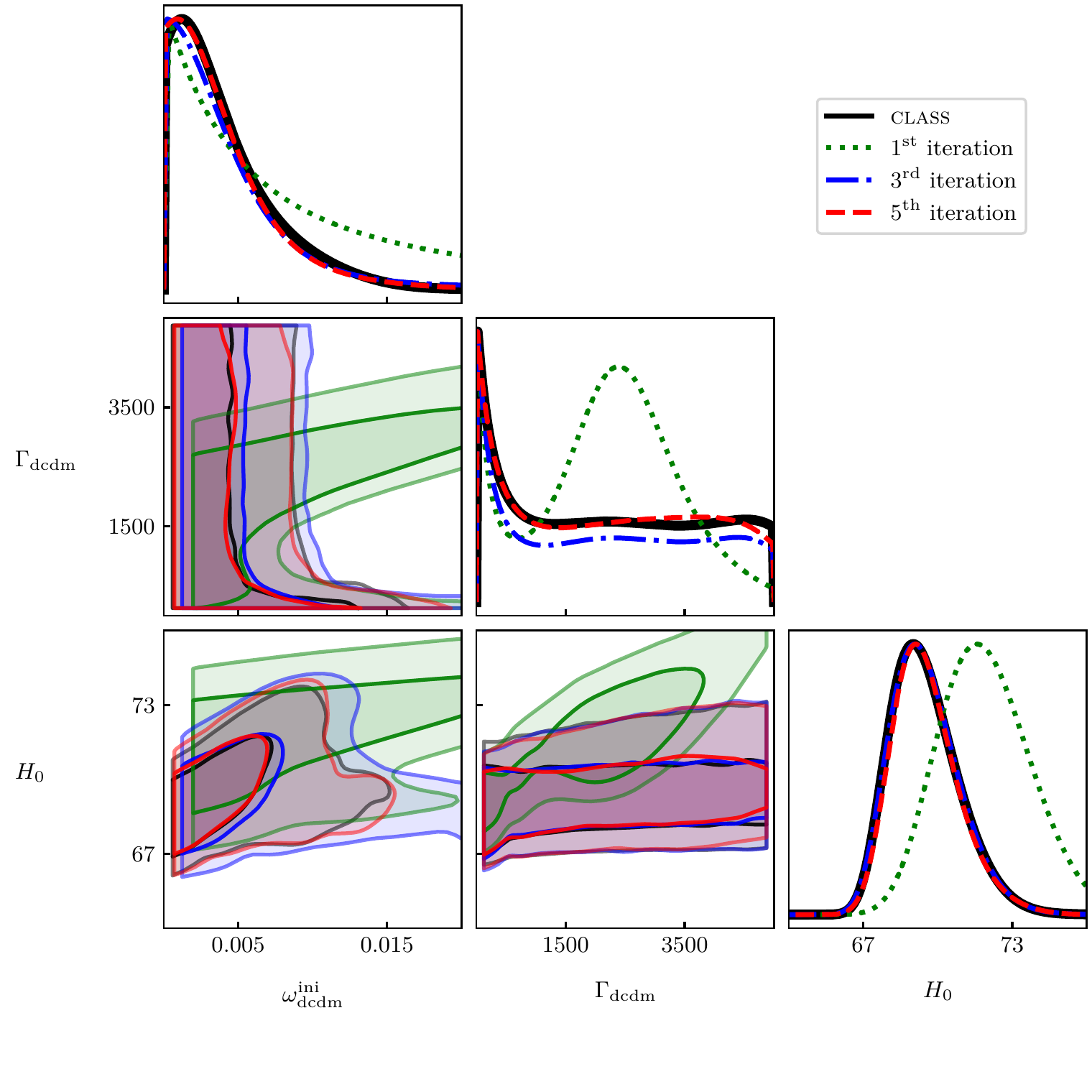}
	\caption{1D and 2D posteriors of the model-specific parameters and $H_0$ for the DCDM model resulting from a standard \class{}-based run (black) and NN models trained on sampled data from iterations 5, 3, and 1 (red, blue, and green).}
	\label{fig:mcmc-dcdm_iterations}
\end{figure}

\begin{figure}[tb]
	\centering
	\includegraphics[width=\textwidth]{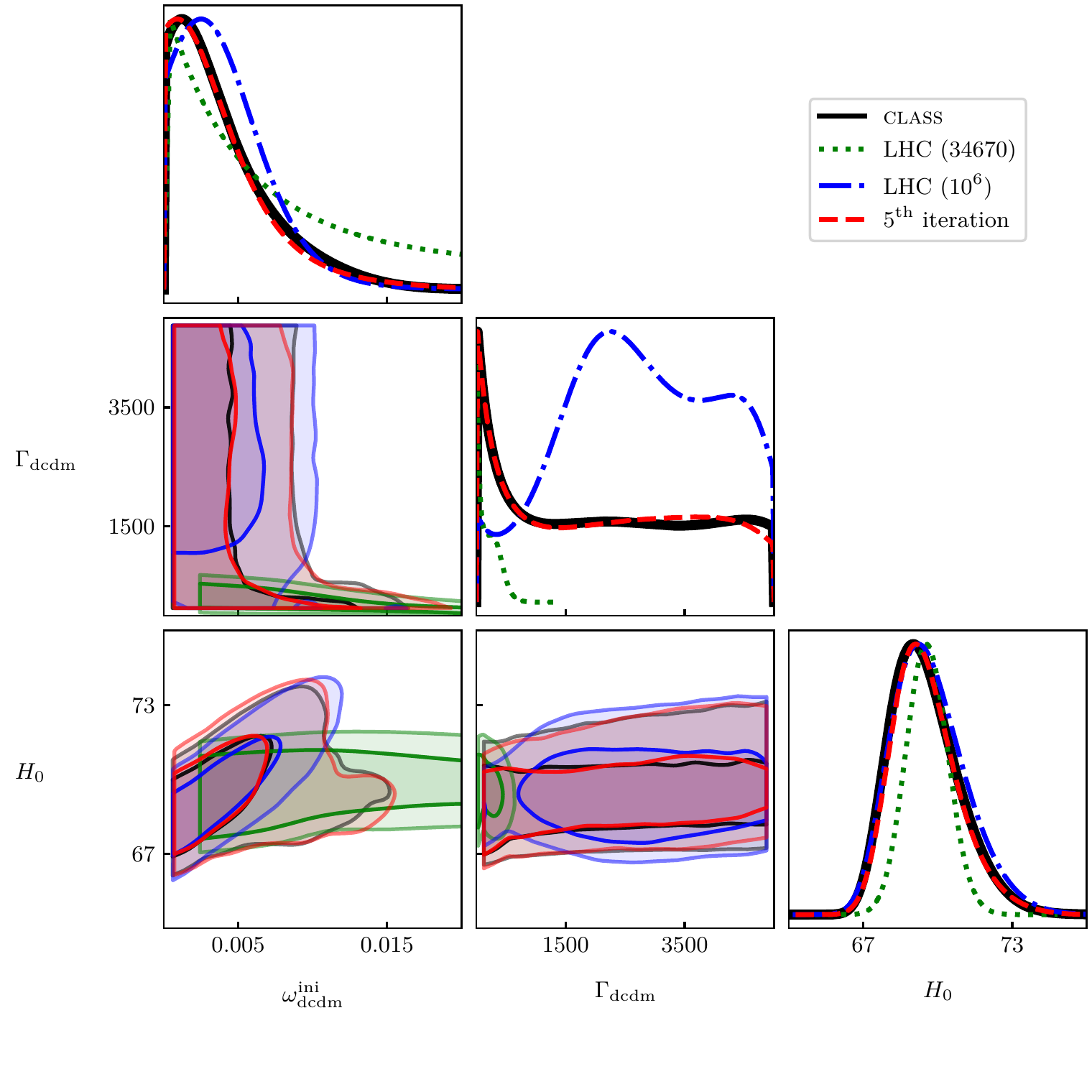}
	\caption{1D and 2D posteriors of the model-specific parameters and $H_0$ for the DCDM model resulting from a standard \class{}-based run (black), an NN model trained on sampled data from iteration 5 (red), and NN models trained on Latin hypercube data with $10^6$ and 34,670 points (blue, green).}
	\label{fig:mcmc-dcdm_lhc}
\end{figure}
Using the plug-in for \montepython{}, we have first sampled the parameter space of the DCDM model which we used to test and validate the training algorithm. In figure~\ref{fig:mcmc-dcdm} we show the posteriors resulting from both a standard \class{}-based MCMC run with approximately $5\times10^5$ accepted chain elements ($\sim 5,000$ CPU core-hours) and a \connect{}-based run with approximately the same number of chain elements ($\sim 10$ CPU core-hours). As can be clearly seen the agreement is excellent! The difference on both means and standard deviations is around $0.01$--$0.1$ standard deviations for all parameters. Figure~\ref{fig:mcmc-dcdm_iterations} shows the \connect{}-based runs using data from iterations 1, 3 and 5 for a subset of the cosmological parameters ($\omega^{\rm ini}_{\rm dcdm}$, $\Gamma_{\rm dcdm}$ and $H_0$), and we can clearly see the progress through the iterations. The first iteration does not lead to a particularly good model, but it manages to find a rough area in which to sample during the following iterations. This improves until the iterations are halted by convergence of the data. When comparing the results of the iterative sampling method to the performance of Latin hypercube sampled NN models in figure~\ref{fig:mcmc-dcdm_lhc}, we see that not even $10^6$ points (individual \class{} computations) are enough for the Latin hypercube sampling to match the results of the iterative sampling, and certainly not a Latin hypercube with as few points as in the dataset from the iterative sampling (34,670 \class{} computations). The Latin hypercubes are of course sampled logarithmically in the $\Gamma_{\rm dcdm}$ parameter as in the iterative case, but this does not help much, the reason being the finite size of the network having to accommodate a huge amount of points in the parameter space that are very far from the region of interest. A network with many more nodes and hidden layers could perhaps be trained to behave nicely in the entire span of the Latin hypercube.


\begin{figure}[tb]
	\centering
	\includegraphics[width=\textwidth]{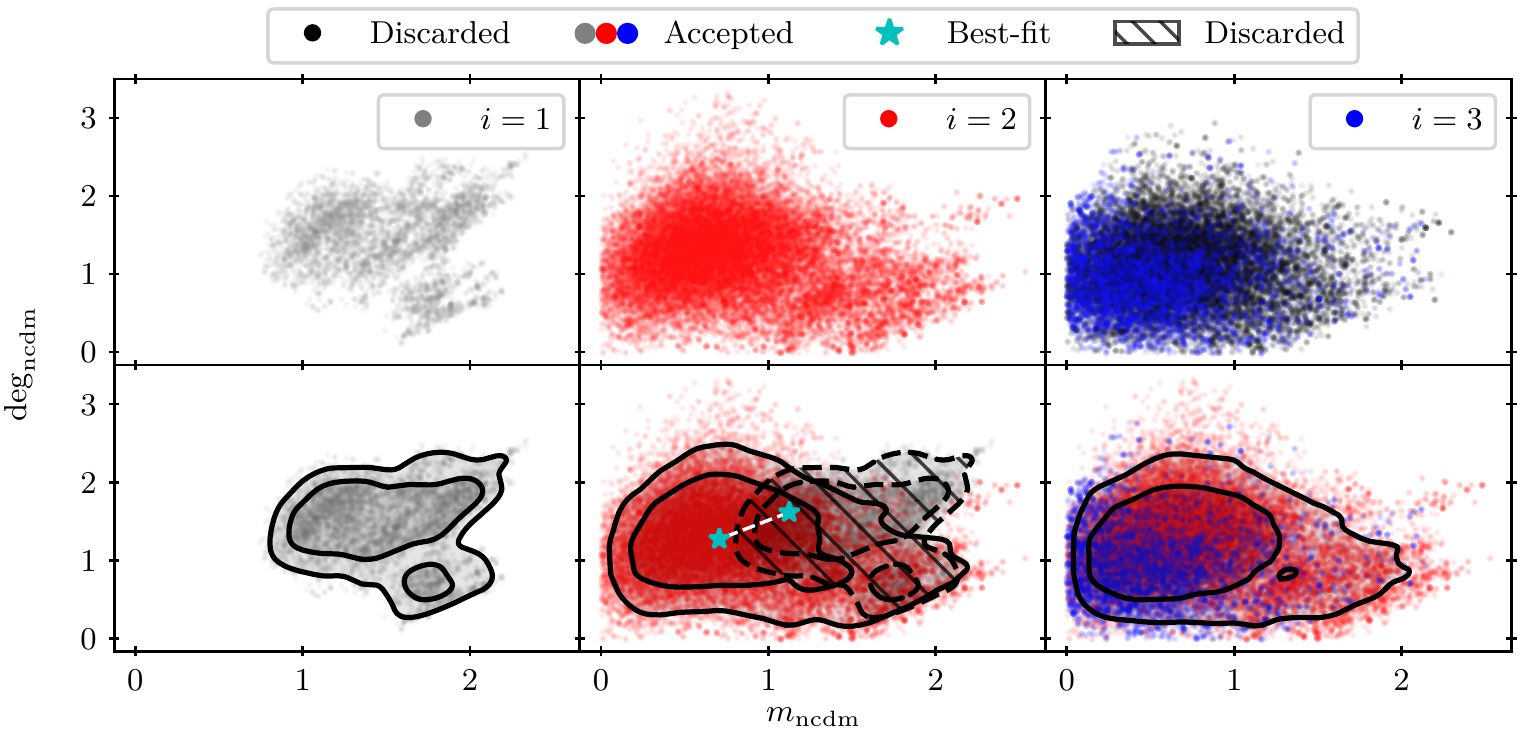}
	\caption{Upper panel: Data from each iteration of a sampling of a cosmological model including massive neutrinos marginalised to the mass-degeneracy plane. Data points from each iteration are filtered to prevent oversampling of certain regions, and the accepted points are shown in colour.
	Lower panel: Combined data after filtering including $1\sigma$ and $2\sigma$ contours. The first iteration is again removed from the dataset, since a better accuracy is achieved without it.}
	\label{fig:iterations_mas}
\end{figure}

\begin{figure}[tb]
	\centering
	\includegraphics[width=\textwidth]{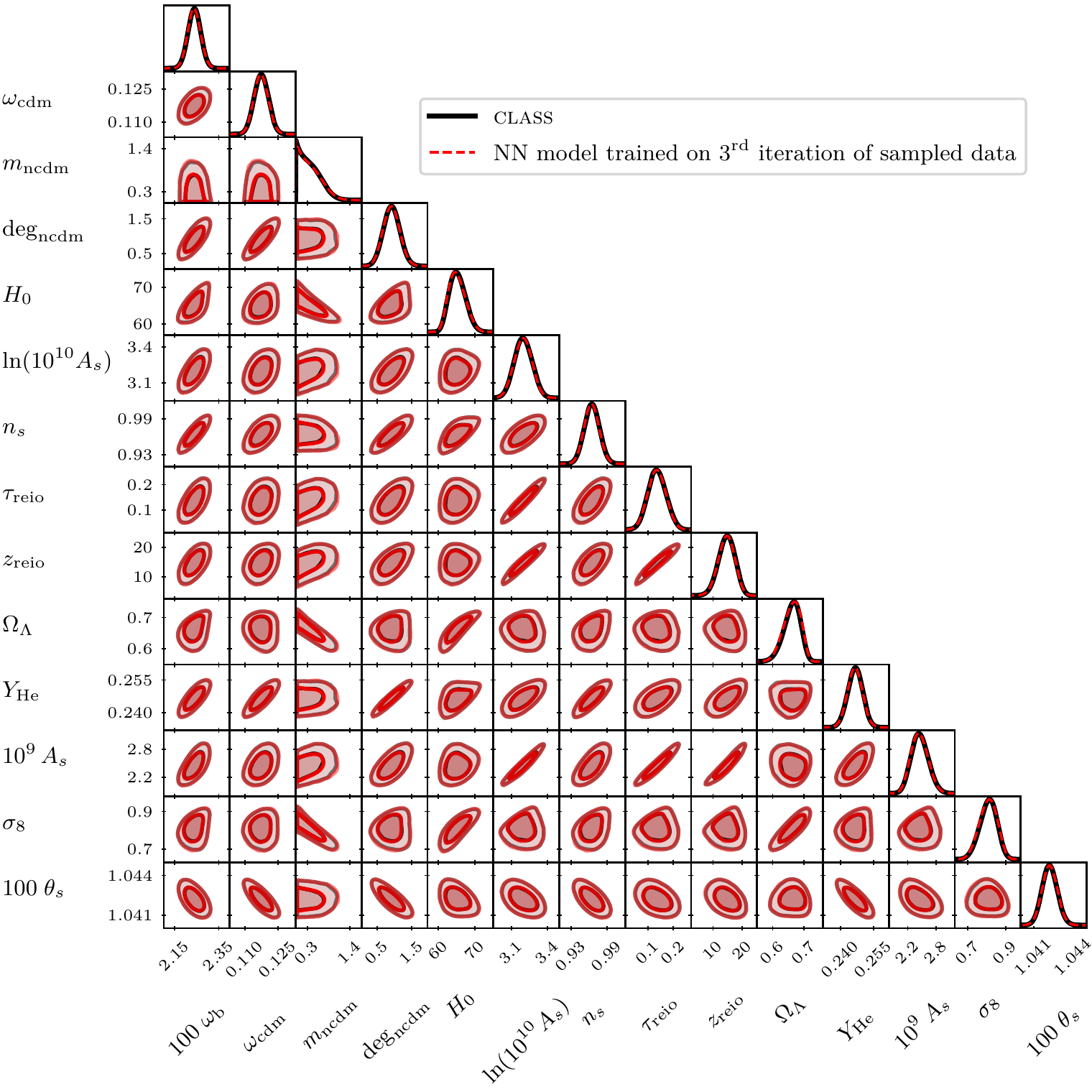}
	\caption{1D and 2D posteriors for the massive neutrino model resulting from a standard \class{}-based run (black) and \connect{} (red).}
	\label{fig:mcmc-mas}
\end{figure}

Next, we have used the exact same setup for a completely different massive neutrino model, described by parameters $m_{\rm ncdm}$ and ${\rm deg}_{\rm ncdm}$. In figure~\ref{fig:iterations_mas} we show the iterative acquisition of training data in the mass-degeneracy plane. As can clearly be seen, this model is significantly easier for the algorithm and converges in just 3 iterations because the likelihood function is significantly more Gaussian than that of the DCDM model. The first iteration again contains 5,000 \class{} computations and the maximal amount of new points from each iteration is now 30,000 (see section~\ref{sec:conclusion} for a discussion hereof). The final dataset contains 30,000 points from $i=2$ and 8,959 points from $i=3$. We then perform the same comparison between MCMC runs with \class{} and \connect{} as in the DCDM case, and the result is shown in figure~\ref{fig:mcmc-mas}. Clearly, in this case the agreement is even better than in the DCDM case, with means and confidence regions differing by around or less than $10^{-2}$ standard deviations for all parameters except $H_0$, which differs by around $10^{-1}$.

\begin{figure}[tb]
	\centering
	\includegraphics[width=\textwidth]{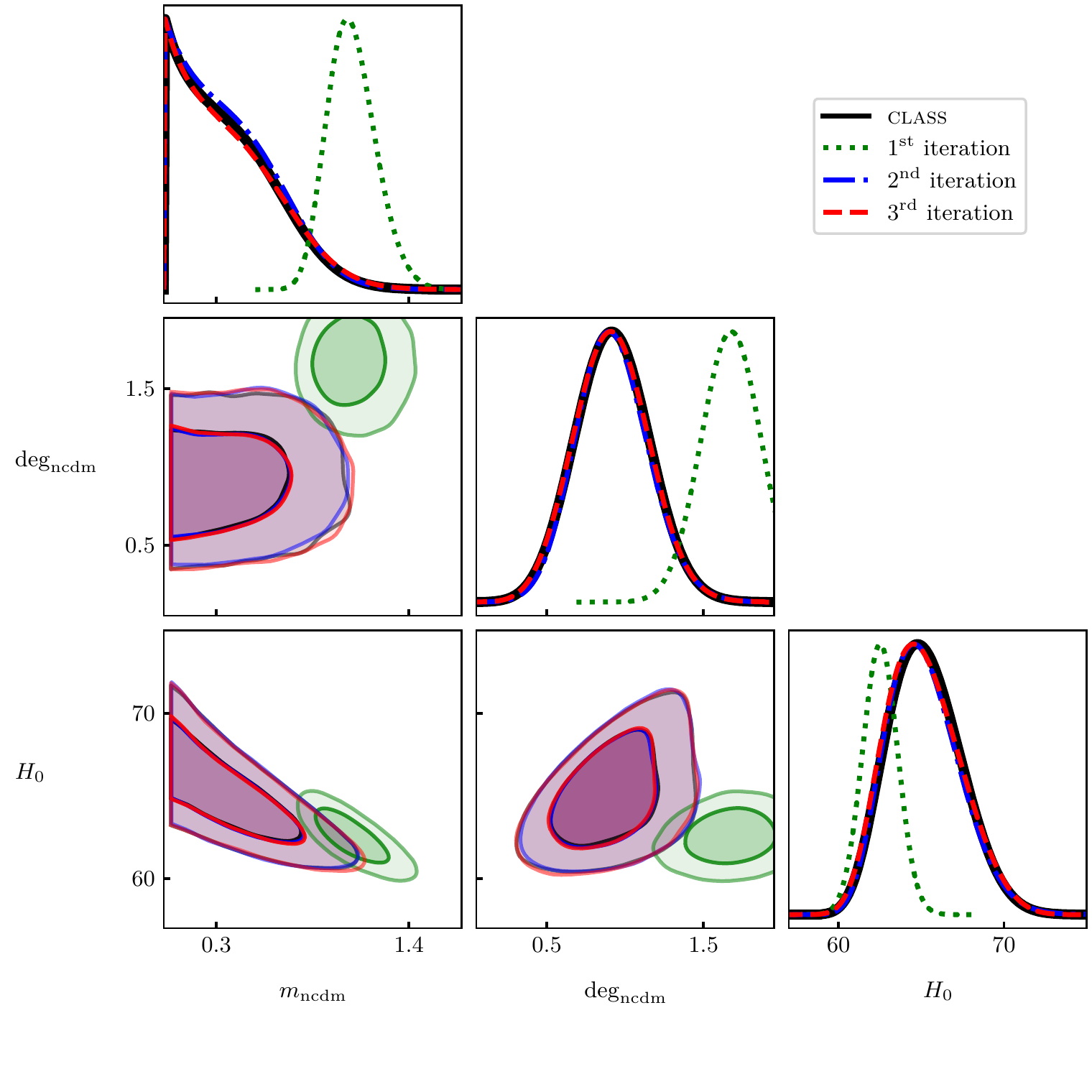}
	\caption{1D and 2D posteriors of the model-specific parameters and $H_0$ for the massive neutrino model resulting from a standard \class{}-based run (black) and NN models trained on sampled data from iterations 3, 2, and 1 (red, blue, and green).}
	\label{fig:mcmc-mas_iterations}
\end{figure}

\begin{figure}[tb]
	\centering
	\includegraphics[width=\textwidth]{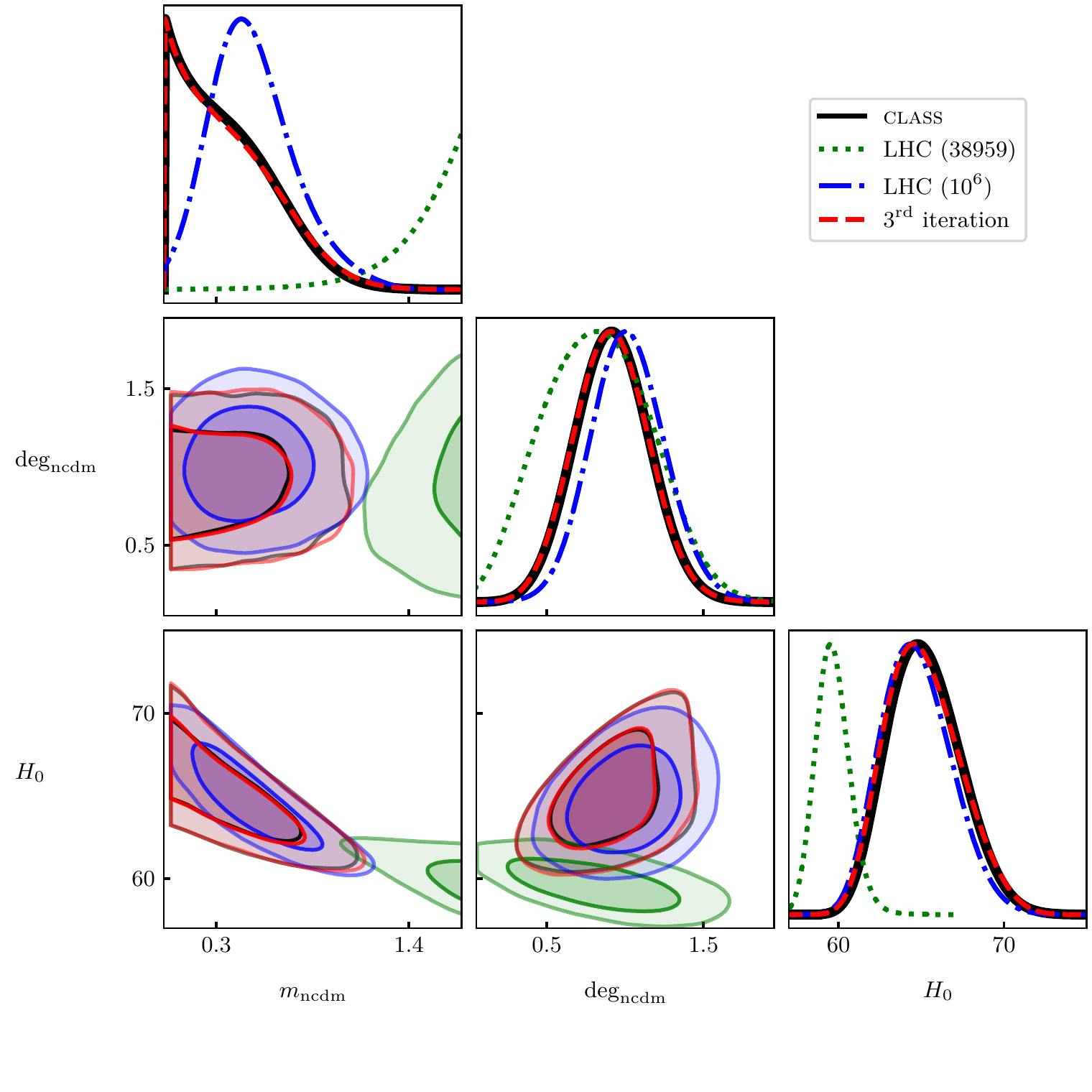}
	\caption{1D and 2D posteriors of the model-specific parameters and $H_0$ for the massive neutrino model resulting from a standard \class{}-based run (black), an NN model trained on sampled data from iteration 3 (red), and NN models trained on Latin hypercube data with $10^6$ and 34,670 points (blue, green).}
	\label{fig:mcmc-mas_lhc}
\end{figure}

Figure~\ref{fig:mcmc-mas_iterations} shows the results of MCMC runs based on the data from the three iterations for a subset of the cosmological parameters ($m_{\rm ncdm}$, ${\rm deg}_{\rm ncdm}$ and $H_0$), and we clearly see that convergence is very quickly achieved since the second iteration is very close to the third. This is quite remarkable due to the fact that the first iteration samples no training data in the immediate vicinity of the best-fit point, so the second iteration being that good really demonstrates how well the sampling works when the likelihood function is simpler and more Gaussian. Figure~\ref{fig:mcmc-mas_lhc} shows how the results of MCMC runs using models trained on Latin hypercubes of different sizes stack up against the results from the last iteration. It is apparent that a Latin hypercube of the same size as the dataset from the iterative process (38,959 points) in no way comes even remotely close to the performance of the last iteration. We also see that a number of points larger than $10^6$ is needed to even come close to the same result as the iterative process, but a much larger dataset would require a larger network architecture as well to perform reasonable, and this would raise the evaluation time for each model and is thus not a viable solution -- not to mention the huge number of \class{} computations that would make the whole idea of emulation obsolete. The larger Latin hypercube seems reasonable for the parameters with a Gaussian posterior, but in the case of a parameter whose likelihood function increases towards a boundary of the prior, we again conclude that the Latin hypercube does not have the ability to represent the cosmological models close to the boundary.

\begin{figure}[tb]
	\centering
	\includegraphics[width=\textwidth]{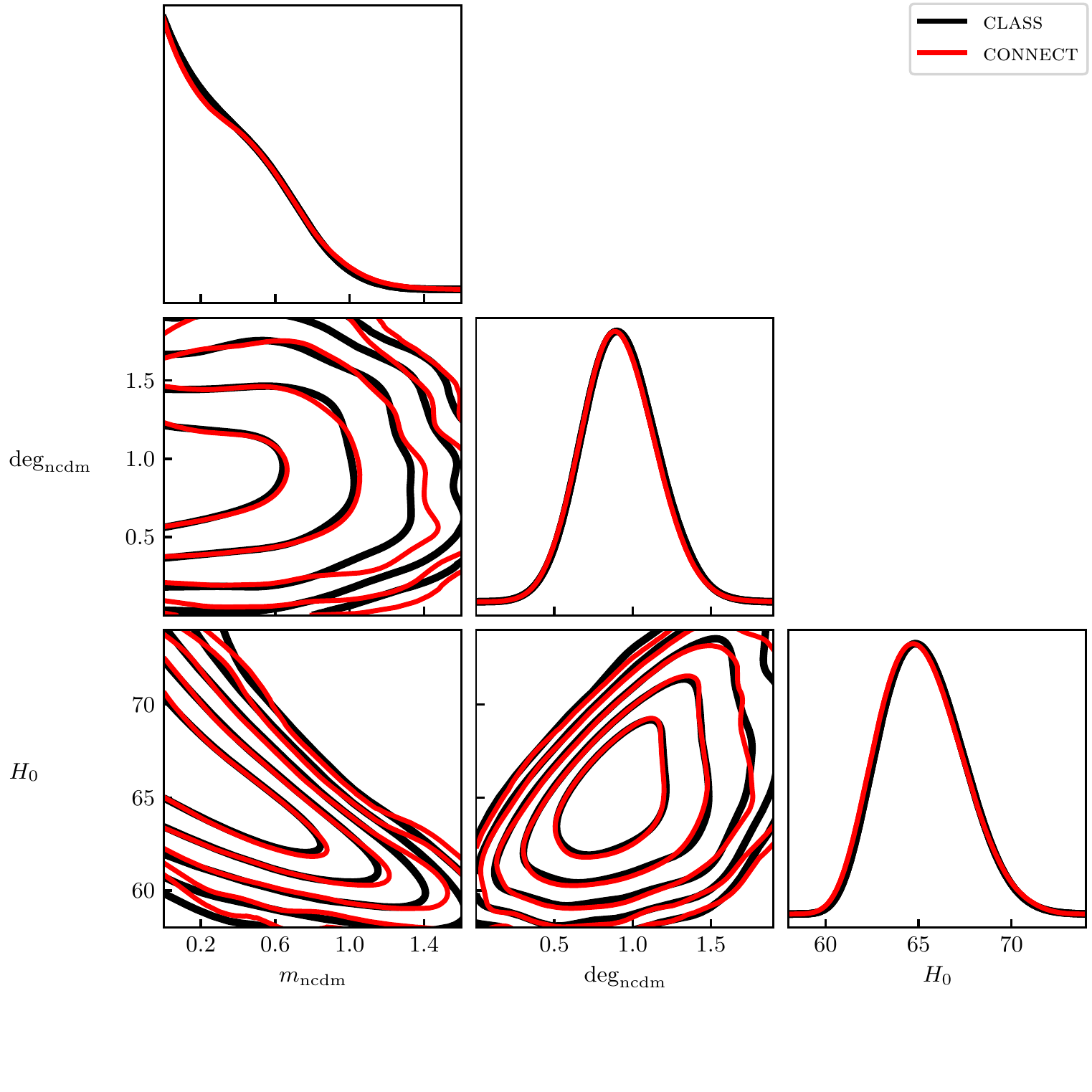}
	\caption{Triangle plot of the marginalised posteriors of selected parameters in the massive neutrino model, as computed directly by \class{} and by the \connect{} model from iteration 3. The contours represent the contours from $1\sigma$ to $5\sigma$, respectively. There is a fair agreement between \connect{} and \class{} even out to $5\sigma$.}
	\label{fig:5sigma}
\end{figure}
Lastly, we note that models trained by \connect{} are also fairly accurate beyond $2\sigma$. Figure~\ref{fig:5sigma} shows the contours of posteriors in the massive neutrino model as estimated by \connect{} and \class{}, respectively, out to $5\sigma$. Evidently, the agreement is reasonable, even out to 5 standard deviations for the massive neutrino model. This accuracy is a consequence of the increased temperature with which the training data sampling chains are run, and the training temperature can be further tuned manually to accommodate even stricter accuracy requirements on a broader region.

\section{Discussion and Conclusions} \label{sec:conclusion}

We have presented \connect{}, a novel, neural network based framework for cosmological parameter inference. The method relies on an iterative MCMC-based generation of training data which proves highly efficient in training the network to perform extremely well for MCMC based cosmological parameter inference. We have tested the robustness and versatility of the method by first building the network structure and training algorithm on a decaying cold dark matter (DCDM) cosmological model and demonstrating that we can achieve results almost exactly identical to well-converged \class{} based MCMC runs, and subsequently used \connect{} in an off-the-shelf manner to run parameter inference on a {\it different} massive neutrino cosmology. Even though the method was not previously tested or optimised on this model we find results which are just as impressive as for the DCDM model.\\

\noindent {\bf Hyperparameters of the iterative sampling.}
A few things can be customised when using the iterative sampling method, including convergence criteria, the size of the initial Latin hypercube, prior bounds for the individual high-temperature MCMC samplings, the maximum number of points to include from each iteration, and the filtration of new data points. The convergence criterion for the individual MCMC runs is chosen to be when $R-1<0.01$ between the chains is valid for all sampling parameters, and the criterion for halting the iterations is chosen to be when the variance between the data acquired from two consecutive iterations is also at $R-1<0.01$. Depending on the complexity of the likelihood, these criteria could be loosened. When filtrating points from each iteration to avoid oversampling in certain regions, the amount of points actually included, of course, decreases over time, and so another halting criterion for the iterations could be when (almost) no new points are added to the total dataset. These two criteria could even be combined, thus giving larger robustness and a better guarantee of the whole best-fit region being sampled. The maximum amount of points included by each iteration is the number of points extracted from the high-temperature MCMC samplings before doing any filtration. This number can be varied depending on the difficulty in sampling the likelihood function. A simpler likelihood function allows for very quick convergence when having a larger maximum amount, which is apparent from our results for massive neutrino model, where this parameter was set to $3\times10^4$ while it was set to $2\times10^4$ for the DCDM results. There can, however, also be a reason to raise this number for more complicated models, since we would then have a better representation from the MCMC chains. This is not ideal for the first few iterations though, since we would waste many more \class{} computations given the low accuracy of the first iterations. The size of the initial Latin hypercube was chosen to $10^4$ points for both of our cosmological models, but this could perhaps be brought down due to the fact that the bounds are chosen to be quite large in order for it to encapsulate all of the regions of significant likelihood. If one knows roughly where the best-fit region is, a much smaller Latin hypercube could be sufficient. The priors on the high-temperature MCMC samplings were chosen to be the same as the bounds of the initial Latin hypercube, but if the hypercube is shrunk, the priors should be set differently than the bounds so as to not exclude significant parts of the parameter space. Due to the nature of the iterative method, convergence should be reached even if the initial Latin hypercube has little to no overlap with the best-fit region (it might take more iterations though), but more tests are needed regarding this.\\

\noindent  {\bf CPU time comparisons.} 
Comparing the CPU intensity of inference run with \connect{} and with standard \class{} based methods is somewhat involved. There is an initial overhead in training the \connect{} emulator for a given model of order $5\times10^4$ \class{} evaluations depending on the number of iterations and how many points each iteration should contribute. However, once the network has been trained the CPU consumption from the \connect{} based \montepython{} inference comes almost exclusively from the likelihood evaluation when using the full Planck likelihood, not from the emulation. A CPU consumption of similar size as evaluations of the neural network arise when using the Planck lite likelihood instead, thus making the speedup much more profound during the MCMC runs. A single evaluation of the $C_\ell$ spectra for a cosmological model takes of order $\SI{5}{\second}$ (depending on the cosmological model -- about twice as much for DCDM) to evaluate using \class{} on a modern intel CPU core while \connect{} only uses around $\SI{3}{\milli\second}$ (including interpolation of the $C_{\ell}$s)! This is an immense speedup of three orders of magnitude, and this is also apparent from the time consumption of an MCMC. The MCMC runs using \class{} as cosmological module each took 200 hours with 6 chains and 6 CPU cores for each chain. This enables the \class{} computations to be parallelised which brings down the total time at the cost of using 6 times the CPU cores. When using \connect{} as cosmological module, we again use 6 chains, but only one CPU core per chain is necessary since the evaluation of the network is not parallelisable. These runs, however, take less than two hours to reach the same level of convergence and a similar amount of accepted steps, so the MCMC runs using \connect{} are sped up by a factor of more than 600. This is in great agreement with the difference in the evaluation times when factoring in the time consumption of the Plank lite likelihood evaluations.

We also need to consider the overhead from sampling of training data, which consists of a number of \class{} computations and, in the case of iterative sampling, high-temperature MCMC sampling and training of neural network models. There is not much to do about the \class{} computations except for limiting the number of them and using many CPU cores on a cluster. With the iterative sampling method the \class{} computations are embarrassingly parallelisable, and the only way to limit the number of computations is to optimise the choice of points in the parameter space to compute and not throw any away in the end. Unfortunately we have to throw away the initial Latin hypercube since the inclusion of this worsens the performance of the network, and for complicated likelihood shapes, we often need to discard the first iteration as well, as argued previously. This means that there are around $10^4$ \class{} computation that we have performed, but cannot use in the final dataset. The high-temperature MCMC samplings are much less expensive in CPU time, but using a normal Metropolis--Hastings algorithm, the sampling is not very parallelisable, and so the time consumption is significant compared to the \class{} computations utilising hundreds of CPU cores at once. The training of the network is quite fast on a GPU, and it normally takes around 5--8 minutes to train our networks using datasets of $\sim 5\times10^4$ points on two GPUs with distributed training, so we can probably not bring this down any further. The computation time of a single iteration in the sampling takes about on hour at this point with 500 CPU cores and two GPU cores allocated. The \class{} computation utilise all CPUs for around 10--15 minutes, whereafter the training uses both GPUs for 5--8 minutes, and lastly the MCMC sampling uses only a handful of CPU cores for around 20--40 minutes depending on the difficulty of achieving convergence. This means that we only use a fraction of the resources for most of the time, but limiting the amount of CPUs to match the MCMC sampling would increase the time of \class{} computations many times. The biggest speedup in the iterations would thus come from the high-temperature MCMC sampling being modified with a more parallelisable algorithm on either a GPU or the many CPU cores available anyway for the \class{} computations. This could bring the sampling part down to a few minutes or even seconds, leaving only \class{} computations and training as the time consuming parts and thus decreasing the time for each iteration to 20 minutes when using the same resources.\\

\noindent {\bf MCMC methods and other sampling strategies.}
For the results presented in this paper we have used the Metropolis--Hastings algorithm for the samplings of parameter space through integration with the popular code \montepython. A benefit of doing this is that we can utilise the entire library of likelihoods contained within \montepython. This means that nothing has to be rewritten and our \connect{} plug-in really provides a plug-and-play solution. The standard Metropolis--Hastings algorithm used for the sampling is, however, not parallelisable to more than a handful of chains running simultaneously, and therefore the MCMC is quite time consuming when compared to everything else during an iteration (assuming enough CPU cores for the \class{} computations to be parallelised). In order to speed up the process, we need to consider alternative sampling methods as well as how to utilise different likelihoods in the analysis. 

There are many ways of sampling the parameter space of a model and MCMC with Metropolis--Hastings is widely chosen mainly due to the cost of evaluating Einstein--Boltzmann solver codes. With the use of neural networks for emulation, a whole new world of parameter inference beyond MCMC sampling opens. Due to the neural network being a smooth function, gradients are not only easy to access but also very numerically stable. This means that gradient-based methods like Hamiltonian Monte Carlo~\cite{Hajian:2006mt} are now a possibility. It is even possible now to move away from MCMC methods and compute profile likelihoods, given that optimisation is so much faster and more robust than when using \class{}. 

In order for us to really utilise the speedup of the emulation, we need some way around the time consumption of likelihood evaluations, which is especially cumbersome for the full Planck likelihood. An idea used by ref.~\cite{SpurioMancini:2021ppk} is to translate the likelihoods into \tf{} syntax in order for it to run rapidly on a GPU. This way one can perform the sampling of the parameter space on the GPU as well, thus having parameter inference in mere seconds (using a parallelisable sampling method such as the affine invariant sampling algorithm~\cite{Foreman-Mackey:2012any}). It requires a lot of work to translate the more heavy likelihoods, so a full library of GPU-compatible likelihoods is probably not realisable in the next few years. A simple solution could also be optimisation of the existing likelihood codes (if possible), since many probably have been written knowing that the \class{} computations will always be much slower and therefore had no reason to be written in the most optimal way. This too requires a significant amount of work, and so this is most likely also not happening in the immediate future. A more easy-to-code solution could in fact be emulation of the likelihood functions themselves. This would allow for both a faster sampling with normal MCMC methods, since the evaluation of the likelihood functions would all be on the level of the \connect{} evaluations, and compatibility with GPUs allowing for more sophisticated gradient-based or parallelisable sampling methods leading to reliable parameter inference in seconds.\\

\noindent {\bf Reproducibility.}
We provide the complete \connect{} framework on GitHub for public use available at \url{https://github.com/AarhusCosmology/connect_public}. All the parameter files used for \connect{} are included as well as a brief description of how to use \connect{} on its own and with \montepython{}. Any version of \montepython{} supporting \class{} as cosmological module can be used with \connect{} without any alterations to the source code. We used our own version of \class{} written in \CC{}, and thus named \CLASSpp{}, but any version supporting the wrapper functions \startus \texttt{lensed_cl()} and \texttt{get_current_derived_parameters()} \stopus can be used. \CLASSpp{} along with a forked version of \montepython{} are both available from our GitHub organisation page \url{https://github.com/AarhusCosmology}.

\section*{Acknowledgements}
We acknowledge computing resources from the Centre for Scientific Computing Aarhus (CSCAA). A.N., E.B.H. and T.T. were supported by a research grant (29337) from VILLUM FONDEN. We would like to thank Christian Fidler and Sven Günther for useful discussions regarding sampling methods and oversampling of regions in the parameter space. We also thank the newborn daughter of A.N., Olivia, for her willingness in letting her father finish this paper.

\appendix

\section{Structure of the code}

\begin{figure}[tb]
\begin{forest}
  pic dir tree,
  pic root,
  for tree={
    directory,
  },
  [codes
    [montepython
    ]
    [class
    ]
    [connect
      [connect.py \annotatefile{Main script -- callable with keywords \texttt{create} and \texttt{train}}, file
      ]
      [source
        [custom\_functions.py \annotatefile{Customise loss and activation functions}, file
        ]
        [architecture \annotatefile{Define new architectures beyond a fully connected network}
        ]
        [\ldots
        ]
      ]
      [input  \annotatefile{Parameter files specifying the creation of data and training hereon.}
      ]
      [trained\_models 
      \annotatefile{Saved models trained by \connect{}}
      ]
      [data  \annotatefile{Training data generated by \connect{}}
        [decaying\_cdm
        ]
        [massive\_nu
        ]
        [\ldots
        ]
       ]
      [mcmc\_plugin \annotatefile{Plug-in for \montepython{} and \cobaya{}}
      ]
    ]
  ]
\end{forest}
\caption{\label{tree:folders} Directory structure of a typical installation. \montepython{}, \class{} and \connect{} are located side-by-side in some root folder \texttt{codes}.}
\end{figure}
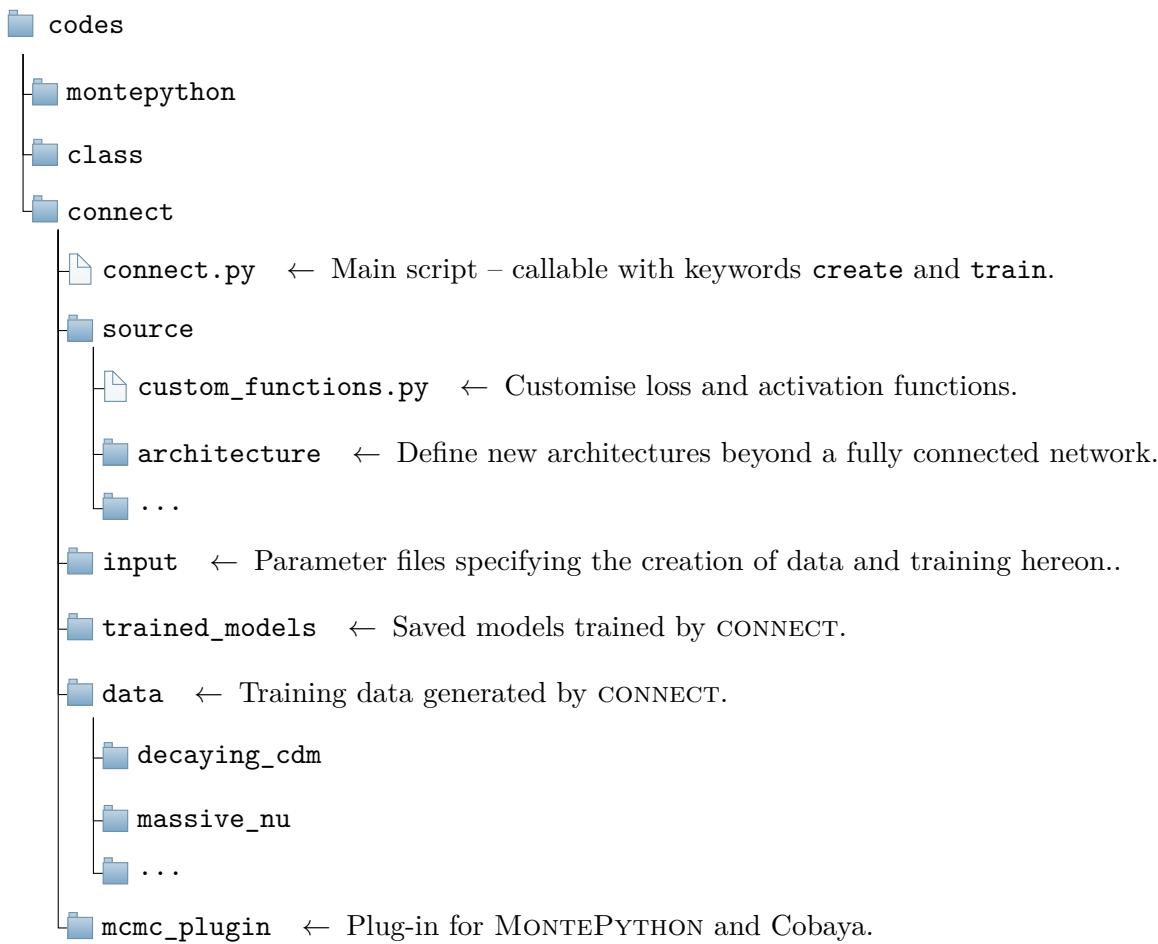

In our implementation we have collected all the classes and functions to be used in a \texttt{source} folder, all the parameter files in an \texttt{input} folder, and all the trained models in a folder called \startus \texttt{trained_models}\stopus. We have also included a space for all the training data, which is automatically sorted in folders with a specified job-name and subfolders with the amount of data points (iteration number) as the name upon creation of the data when using Latin hypercube sampling (iterative sampling). These data folders are collected in the folder \texttt{data}. Within the \texttt{source} folder, the \startus \texttt{custom_functions.py} \stopus file contains classes for adding new activation functions and loss functions, which gives the user an easy way of implementing new custom functions for training networks. Within the \texttt{source/architecture} folder, users can in addition easily define entirely new architectures for the network and tailor everything to their needs. In the folder \startus \texttt{mcmc_plugin} \stopus we have made a wrapper for trained models to mimic \class{}. This structure is depicted in figure~\ref{tree:folders}.

\bibliographystyle{utcaps}
\bibliography{connect2022}

\end{document}